\RequirePackage{fix-cm}
\documentclass[twocolumn,epjc3]{svjour3}
\smartqed  
\RequirePackage{graphicx}
\usepackage{hyperref}
\usepackage{fouridx}
\usepackage{epstopdf}
\usepackage{amsmath}
\usepackage{hyperref}

\journalname{Eur. Phys. J. C}

\begin{document}

\title{Angular distributions in the radiative decays of the \boldmath{${}^3D_3$} state of charmonium originating from polarized \boldmath{$\bar{p}p$} collisions}


\author{Cheuk-Ping Wong\thanksref{e1,addr1}
        \and
        Alex W. K. Mok\thanksref{e2,addr2}
        \and
        Wai-Yu Sit\thanksref{e3, addr2}
}

\thankstext{e1}{e-mail: cwong14@student.gsu.edu}
\thankstext{e2}{e-mail: wkmok@hkbu.edu.hk}
\thankstext{e3}{e-mail: 12466654@hkbu.edu.hk}

\institute{Department of Physics and Astronomy, Georgia State University, Atlanta, GA 30303, USA\label{addr1}
          \and
          Department of Physics, Hong Kong Baptist University, Kowloon Tong, Hong Kong\label{addr2}
}

\date{Received: date / Accepted: date}

\maketitle

\begin{abstract}
Using the helicity formalism, we calculate the combined angular distribution function of the two gamma photons ($\gamma_1$ and $\gamma_2$) and the electron ($e^-$) in the triple cascade process $\bar{p}p\rightarrow{}^3D_3\rightarrow{}^3P_2+\gamma_1\rightarrow(\psi+\gamma_2) +\gamma_1 \rightarrow (e^- + e^+) +\gamma_2 +\gamma_1$, when $\bar{p}$ and $p$ are arbitrarily polarized. We also derive six different partially integrated angular distribution functions which give the angular distributions of one or two particles in the final state. Our results show that by measuring the two-particle  angular distribution of $\gamma_1$ and $\gamma_2$ and that of $\gamma_2$ and $e^-$, one can determine the relative magnitudes as well as the relative phases of all the helicity amplitudes in the two charmonium radiative transitions ${}^3D_3\rightarrow{}^3P_2+\gamma_1$ and $^3P_2\rightarrow \psi+\gamma_2$.\end{abstract}
\allowdisplaybreaks
\section{Introduction}
\label{sec:Intro}
The study of charmonium states above the open charm $D\bar{D}$ threshold of 3.73 GeV has captured much attention in the theoretical and experimental community recently \cite{HigherCharmonia,NewStatesAboveCharmThreshold,QuarkoniaAndTheirTransitions,CharmoniumSpectroscopyAboveThresholds,ProgressPuzzlesAndOpportunities,ExcitedAndExoticCharmoniumSpectroscopyFromLatticeQCD,CharmoniumAndBottomoniumInHeavyIonCollisions}. Among the higher charmonium states, the unobserved ${}^3D_3$ state is quite interesting as its decay width is expected to be narrow. Although the strong decay of the ${}^3D_3$ state to $D\bar{D}$ is Zweig-allowed, it is suppressed by the \textit{F}-wave centrifugal barrier factor. This dominant decay width is predicted to be less than 1 MeV \cite{Charmonium options for the X(3872),B-meson gateways to missing charmonium levels,Charmonium levels near threshold and the narrow state} and thus the radiative transition of ${}^3D_3 \rightarrow \gamma + {}^3P_2$ may be observable \cite{QuarkoniaAndTheirTransitions,CharmoniumSpectroscopyAboveThresholds}. The measurement of the angular distributions in the radiative decay of this charmonium state can provide valuable information on the true dynamics of the charmonium system above the charm threshold. In fact, charminoum spectroscopy is a key element of the planned $\bar{\text{P}}$ANDA experiments at GSI \cite{The PANDA experiment at FAIR,Physics performance report for PANDA: strong interaction studies with antiprotons}, which will carry out systematic high-precision study of charmonium states below and above the charm threshold in $\bar{p}p$ annihilation.

In our previous paper \cite{Our paper}, it is shown that by measuring the joint angular distribution of the two photons ($\gamma_1,\gamma_2$) and that of the second photon and electron ($\gamma_2, e^-$), in the sequential decay process originating from unpolarized $\bar{p}p$ collisions, namely, $\bar{p}p\rightarrow{}^3D_3\rightarrow{}^3P_2+\gamma_1\rightarrow (\psi+\gamma_2) +\gamma_1 \rightarrow (e^- + e^+) +\gamma_2 +\gamma_1$, one can extract the relative magnitudes as well as the cosines of the relatives phases of all the angular-momentum helicity amplitudes in the radiative decay processes ${}^3D_3 \rightarrow {}^3P_2 + \gamma_1$ and ${}^3P_2 \rightarrow \psi +\gamma_2$. The sines of the relative phases of these helicity amplitudes, however, cannot be determined uniquely. By considering the sequential decay of ${}^3D_3$ produced in polarized $\bar{p}p$ collisions, one may also obtain unambiguously the sines of the relative phases. So in this paper we calculate the angular distributions of the final stable decay products, $\gamma_1, \gamma_2$ and $e^-$, in the above cascade process when both $\bar{p}$ and $p$ are arbitrarily polarized. Our final model-independent expressions for the angular distribution functions are valid in the $\bar{p}p$ center-of-mass frame and they are written as sums of terms involving products of the Wigner $D$-functions whose arguments are the angles representing the directions of the final electron and of the two photons. The coefficients in these expansions are functions of the angular-momentum helicity amplitudes which contain all the dynamics of the individual decay processes. They are also functions of the longitudinal and the transverse components of the polarization vector of $\bar{p}$ and $p$ in their respective rest frames.

Potential model calculations show that the helicity amplitudes are in general complex \cite{ref 11} and thus their relative phases are nontrivial. Once the angular distributions in polarized $\bar{p}p$ collisions are experimentally measured, our expressions will enable one to determine the relative magnitudes as well as the relative phases of all the complex angular-momentum helicity amplitudes in the radiative decay processes ${}^3D_3 \rightarrow {}^3P_2 + \gamma_1$ and ${}^3P_2 \rightarrow \psi +\gamma_2$. It is important that both $\bar{p}$ and $p$ are polarized to get this complete information. We will derive the angular distribution functions by means of density matrix formalism where the density matrix elements are given in terms of the polarization vectors defined for stationary antiproton and proton. Our results are valid even when $\bar{p}$ and $p$ have arbitrary momenta since the density matrix elements are Lorentz invariant \cite{ref 12}.

The format of the rest of the paper is as follows: In Sect. 2, we give the calculation for the combined angular distribution function of the electron and of the two photons in the cascade process $\bar{p}p\rightarrow{}^3D_3\rightarrow{}^3P_2+\gamma_1\rightarrow( \psi+\gamma_2) +\gamma_1 \rightarrow (e^- + e^+) +\gamma_2 +\gamma_1$, when $\bar{p}$ and $p$ are arbitrarily polarized. We then show how the measurement of this combined angular distribution of $\gamma_1$, $\gamma_2$ and $e^-$ enables us to obtain complete information on the helicity amplitudes in the two radiative transitions ${}^3D_3 \rightarrow {}^3P_2 + \gamma_1$ and ${}^3P_2 \rightarrow \psi +\gamma_2$. In Sect. 3, we present the results for the partially integrated angular distributions in six different cases where the combined angular distribution function of the three particles is integrated over the directions of one or two particles. We also show how the measurement of these simpler angular distributions will again give all the information there is to get on the helicity amplitudes. Finally, in Sect. 4, we make some concluding remarks.

\section{The combined angular distribution function of the photons and electron}
\label{sec:2}
\begin{figure}
 \includegraphics[scale=0.4]{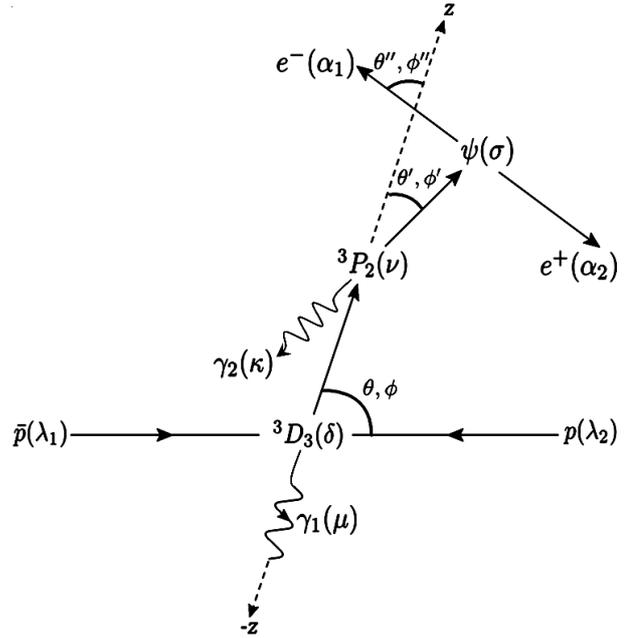}
\caption{Symbolic sketch of   $\bar{p}(\lambda_1)+p(\lambda_2)\rightarrow\fourIdx{3}{}{}{3}{D}(\delta)\rightarrow\fourIdx{3}{}{}{2}{P}(\nu)+\gamma_1(\mu)
  \rightarrow[\psi(\sigma)+\gamma_2(\kappa)]+\gamma_1(\mu)\rightarrow[e^-(\alpha_1)+e^+(\alpha_2)]
  +\gamma_1(\mu)+\gamma_2(\kappa)$ showing different angles of the decay particles.}
\label{graph}       
\end{figure}

We consider the cascade process, $\bar{p}(\lambda_1)+p(\lambda_2)\rightarrow{}^3D_3(\delta)$ $\rightarrow{} ^3P_2(\nu)+\gamma_1(\mu)\rightarrow[\psi(\sigma)+\gamma_2(\kappa)]+\gamma_1(\mu)\rightarrow[e^-(\alpha_1)+e^+(\alpha_2)]+\gamma_2(\kappa)+\gamma_1(\mu)$, in the $^3D_3$ rest frame or the $\bar{p}p$ c.m. frame. The Greek symbols in the brackets represent the helicities of the particles except $\delta$ which represents the $z$ component of the angular momentum of the stationary $^3D_3$ resonance. We choose the $z$ axis to be the direction of motion of $^3P_2$ in the $^3D_3$ rest frame. The $x$ and $y$ axes are arbitrary and the experimentalists can choose them according to their convenience. A symbolic sketch of the cascade process is shown in Fig.~\ref{graph}.

Following the conventions of our previous paper \cite{Our paper}, the probability amplitude for the cascade process can be expressed in terms of the Wigner $D$-function as
\begin{align}
T^{\alpha_1\alpha_2\mu\kappa}_{\lambda_1\lambda_2}&=\frac{7\sqrt{15}}{16\pi^2}
	B_{\lambda_1\lambda_2}C_{\alpha_1\alpha_2}\sum^{-3\rightarrow3}_{\delta}\sum^{-1\rightarrow1}_{\sigma}
	A_{\mu+\delta,\mu}E_{\sigma\kappa}\nonumber\\
	&\quad\times D^3_{\delta\lambda}(\phi,\theta,-\phi)D^{2*}_{\mu+\delta,\sigma-\kappa}(\phi',\theta',-\phi')\nonumber\\
	&\quad\times D^{1*}_{\sigma\alpha}(\phi'',\theta'',-\phi'')\text{,}
\label{eq:PA}
\end{align}
where $B_{\lambda_1\lambda_2}$,$A_{\nu\mu}$, $E_{\sigma,\kappa}$ and $C_{\alpha_1\alpha_2}$ are the angular momentum helicity amplitudes for the individual sequential processes $\bar{p}p\rightarrow{}^3D_3$, $^3D_3\rightarrow{}^3P_2+\gamma_1$, $^3P_2\rightarrow\psi+\gamma_2$ and $\psi\rightarrow e^++e^-$, respectively. In the $D$-functions, the angles $(\phi,\theta)$ giving the direction of $\bar{p}$, the angles $(\phi',\theta')$ giving the direction of $\psi$ and the angles $(\phi'',\theta'')$ giving the direction of $e^-$ are measured in the $^3D_3$, the $^3P_2$ and the $\psi$ rest frames, respectively. The angles of each decay particle observed in different rest frames can be calculated using the Lorentz transformation. The equations relating these angles are given in \cite{Mok 2008}.

Because of the $C$ and $P$ invariances \cite{Elementary Particle Theory}, the angular momentum helicity amplitudes in (\ref{eq:PA}) are not all independent. We have
\begin{align}
B_{\lambda_1\lambda_2}&\overset{P}{=}B_{-\lambda_1-\lambda_2}\text{,}\nonumber\\
B_{\lambda_1\lambda_2}&\overset{C}{=}B_{\lambda_2\lambda_1}\text{,}\nonumber\\
A_{\nu\mu}&\overset{P}{=}A_{-\nu-\mu}\text{,}\nonumber\\
E_{\sigma\kappa}&\overset{P}{=}E_{-\sigma-\kappa}\text{,}\nonumber\\
C_{\alpha_1\alpha_2}&\overset{P}{=}C_{-\alpha_1-\alpha_2}\text{,}\nonumber\\
C_{\alpha_1\alpha_2}&\overset{C}{=}C_{\alpha_2\alpha_1}\text{.}
\label{symmetry relation}
\end{align}
Making use of the symmetry relations of (\ref{symmetry relation}), we now re-label the independent angular-momentum helicity amplitudes as follows:
\begin{alignat}{2}
&B_0=\sqrt{2}B_{\frac{1}{2}\frac{1}{2}}\text{,}&\quad&B_1=\sqrt{2}B_{\frac{1}{2}-\frac{1}{2}}\text{,}\nonumber\\
&A_i=A_{i-2,1}=A_{2-i,-1}&&(i=0,1,2,3,4)\text{,}\nonumber\\
&E_j=E_{j-1,1}=E_{1-j,-1}&&(j=0,1,2)\text{,}\nonumber\\
&C_0=\sqrt{2}C_{\frac{1}{2}\frac{1}{2}}\text{,}&&C_1=\sqrt{2}C_{\frac{1}{2}-\frac{1}{2}}\text{.}
\end{alignat}
We will also make use of the following normalizations:
\begin{eqnarray}\label{eq: normalizations 1}
|B_0|^2+|B_1|^2=|C_0|^2+|C_1|^2=1
\end{eqnarray}
\noindent
and
\begin{eqnarray}\label{eq: normalizations 2}
\sum^{4}_{i=0}|A_i|^2=\sum^{2}_{j=0}|E_j|^2=1\text{.}
\end{eqnarray}

The normalized angular distribution function for the cascade process when the initial $\bar{p}$ and $p$ are arbitrary polarized and the final polarizations of $\gamma_1$, $\gamma_2$, $e^-$ and $e^+$ are not observed is given by
\begin{align}
&W(\theta,\phi;\theta',\phi';\theta'',\phi'')\nonumber\\
	&=N\sum^{\pm\frac{1}{2}}_{\lambda_1\lambda_2\lambda'_1\lambda'_2}\sum^{\pm\frac{1}{2}}_{\alpha_1\alpha_2}\sum^{\pm1}_{\mu\kappa}
		T^{\alpha_1\alpha_2,\mu\kappa}_{\lambda_1\lambda_2}\rho_{1_{\lambda_1\lambda'_1}}\rho_{2_{\lambda_2\lambda'_2}}
		T^{\alpha_1\alpha_2,\mu\kappa*}_{\lambda'_1\lambda'_2}\text{,}
\label{eq:Normalized angular distribution}
\end{align}
where $N$ is the normalization constant. It is determined by requiring that for the unpolarized case the integral of the angular distribution function $W(\theta,\phi;\theta',\phi';\theta'',\phi'')$ over all the directions of $\gamma_1$, $\gamma_2$ and $e^-$ or over all the angles, $(\theta,\phi;\theta',\phi';\theta'',\phi'')$, is $1$. In (\ref{eq:Normalized angular distribution}) the symbols $\rho_{1_{\lambda_1\lambda'_1}}$ and $\rho_{2_{\lambda_2\lambda'_2}}$ represent the density matrices of $\bar{p}$ and $p$, respectively. In the helicity basis states of the particles these matrix elements are \cite{ref 15}
\begin{align}
\rho_{1_{\lambda_1\lambda'_1}}
=\chi^\dagger_{\lambda_1}\frac{1}{2}(1+\vec{P_1}\cdot\vec{\sigma})\chi_{\lambda'_1}
\label{eq:Rho 1 matrix}
\end{align}
and
\begin{align}
\rho_{2_{\lambda_2\lambda'_2}}
=\beta^\dagger_{\lambda_2}\frac{1}{2}(1+\vec{P_2}\cdot\vec{\sigma})\beta_{\lambda'_2}\textbf{,}
\label{eq:Rho 2 matrix}
\end{align}
where $\vec{\sigma}$ are the Pauli matrices. In (\ref{eq:Rho 1 matrix}) and (\ref{eq:Rho 2 matrix}) $\vec{P_{1}}$ and $\vec{P_{2}}$ are the polarization vectors of $\bar{p}$ and $p$ and the two-component helicity eigenstates $\chi_{\lambda_1}$ of $\bar{p}$ and $\beta_{\lambda_2}$ of $p$ satisfy
\begin{align}
\vec{\sigma}\cdot\hat{p}\chi_{\lambda_1}=\lambda_1\chi_{\lambda_1}
\end{align}
and
\begin{align}
\vec{\sigma}\cdot(-\hat{p})\beta_{\lambda_2}=\lambda_2\beta_{\lambda_2}\textbf{,}
\end{align}
where $\hat{p}$ is the direction of the momentum of $\bar{p}$ and $\lambda_1$ and $\lambda_2$ can take the values $+1$ or $-1$. In the coordinate system we defined in the beginning, we have
\begin{eqnarray}
\chi_+=\begin{bmatrix}
		\cos(\theta/2)\\
		\sin(\theta/2)\exp(\text{i}\phi)
		\end{bmatrix}\text{,}
\end{eqnarray}
\noindent
and
\begin{eqnarray}
\chi_-=\begin{bmatrix}
		-\sin(\theta/2)\exp(-\text{i}\phi)\\
		\cos(\theta/2)
		\end{bmatrix}
\end{eqnarray}
and the phase of $\beta$ is such that \cite{Elementary Particle Theory}
\begin{align}
\beta_{\mp}=\chi_{\pm}\text{.}
\end{align}
(\ref{eq:Rho 1 matrix}) can be rewritten as
\begin{align}
\rho_{1_{\lambda_1\lambda'_1}}
&=\chi^\dagger_{\lambda_1}\frac{1}{2}
	\begin{bmatrix}
	1+P_{1z}          & P_{1x}- P_{1y}\\
	P_{1x}+\text{i} P_{1y} & 1-P_{1z}
	\end{bmatrix}\chi_{\lambda'_1}\nonumber\\
&=\frac{1}{2}
	\begin{bmatrix}
	1+P_{1z'}           & P_{1x'}-\text{i} P_{1y'}\\
	P_{1x'}+\text{i} P_{1y'} & 1-P_{1z'}
    \end{bmatrix}\text{,}
\label{eq:rewritten Rho 1 matrix}
\end{align}
where the unit vectors along the new $x'$, $y'$ and $z'$ axes are related to the corresponding vectors of the $xyz$ coordinate system by
\begin{align}
\hat{\text{i}'}&=(\sin^2\phi+\cos\theta\cos^2\phi)\hat{\text{i}}\nonumber\\
	&\quad-(\sin\phi\cos\phi-\cos\theta\sin\phi\cos\phi)\hat{\text{j}}-\cos\phi\sin\theta\hat{\text{k}}\text{,}\nonumber\\
\hat{\text{j}'}&=(-\cos\phi\sin\phi+\cos\theta\cos\phi\sin\phi)\hat{\text{i}} \nonumber\\
	&\quad+(\cos^2\phi+\cos\theta\sin^2\phi)\hat{\text{j}}-\sin\phi\sin\theta\hat{\text{k}}\text{,}\nonumber\\
\hat{\text{k}'}&=\sin\theta\cos\phi\hat{\text{i}}-\sin\theta\sin\phi\hat{\text{j}}+\cos\theta\hat{\text{k}}\text{.}
\end{align}
Similarly, (\ref{eq:Rho 2 matrix}) can be rewritten as
\begin{align}
\rho_{2_{\lambda_2\lambda'_2}}
&=\beta^\dagger_{\lambda_2}\frac{1}{2}
	\begin{bmatrix}
	1+P_{2z}          & P_{2x}-\text{i} P_{2y}\\
	P_{2x}+\text{i} P_{2y} & 1-P_{2z}
	\end{bmatrix}\beta_{\lambda'_2}\nonumber\\
&=\frac{1}{2}
	\begin{bmatrix}
	1-P_{2z'}           & P_{2x'}+\text{i} P_{2y'}\\
	P_{2x'}-\text{i} P_{2y'} & 1+P_{2z'}
	\end{bmatrix}\text{.}
\label{eq:rewritten Rho 2 matrix}
\end{align}
In (\ref{eq:rewritten Rho 1 matrix}) and (\ref{eq:rewritten Rho 2 matrix}), $P_{1z'}$ and $-P_{2z'}$ are the longitudinal components (components along the momenta of the respective particles) and the $x'$ and $y'$ components are the transverse componets of the polarization vectors. Note that $W(\theta,\phi;\theta',\phi';\theta'',\phi'')$ is now given in terms of the density matrix elements defined for stationary proton and antiproton. But (\ref{eq:Normalized angular distribution}) is of course valid in the $\bar{p}p$ c.m. frame, where $\bar{p}$ and $p$ are moving with relativistic velocities, since the density matrix elements are Lorentz invariant \cite{ref 12}.

Substituting (\ref{eq:PA}) into (\ref{eq:Normalized angular distribution}) and performing the various sums will give us a useful expression for the angular distribution function $W(\theta,\phi;\theta',\phi';\theta'',\phi'')$ in terms of the Wigner $D$-functions. Before we do the sums we make use of the Clebsch-Gordan series relation for the $D$-functions, namely,
\begin{align}
&D^{j_1}_{m_1m_2}D^{j_2}_{m'_1m'_2}\nonumber\\
&=\sum^{j_1+j_2}_{J=|j_1-j_2|}
	\langle j_1j_2m_1m'_1|J,m_1+m'_1\rangle\nonumber\\
 	&\quad\times\langle j_1j_2m_2m'_2|J,m_2+m'_2\rangle D^J_{m_1+m'_1,m_2+m'_2}
\end{align}
and the relation
\begin{align}
D^{j*}_{m_1m_2}=(-1)^{m_1-m_2}D^j_{-m_1-m_2}\text{.}
\end{align}
After a long calculation, we obtain 
\begin{align}
\label{eq:W2}
&W(\theta,\phi;\theta',\phi';\theta'',\phi'')\nonumber\\
&=\frac{1}{4(4\pi)^{3}}
	\sum_{J_1}^{0\rightarrow6}\sum_{J_2}^{0\rightarrow4}\sum_{J_3}^{0\rightarrow2}
	\sum_{d}^{0\rightarrow d_m}\sum_{d'}^{0\rightarrow d'_m}\gamma_{J_3}\alpha^{J_1J_2}_d\epsilon^{J_3J_2}_{d'}\nonumber\\
&\quad\times\sum_{M(J_1)}(2-\delta_{M0})
	[\beta^{J_1}_MY^{J_1J_2J_3}_{dd'M}+(-1)^{J_1}\beta^{J^*_1}_MY^{J_1J_2J^*_3}_{dd'M}]\text{,}
\end{align}
where
\begin{align}
  d_m&= \operatorname{min}\{J_1,J_2\}\text{,}\nonumber\\
  d'_m&= \operatorname{min}\{J_2,J_3\}\text{,}\nonumber\\
  M(J_1)&= \left\{
  \begin{array}{l l}
     0\rightarrow J_1 & \quad \text{when $J_1$=0,1,2}\\
     0,1,2 & \quad \text{when $J_1$=3,4,5,6}\\
  \end{array} \right.\text{.}
\end{align}
The angle-dependent function $Y^{J_1J_2J_3}_{dd'M}$ in (\ref{eq:W2}) is defined by
\begin{align}
Y^{J_1J_2J_3}_{dd'M}&=D^{J_3*}_{d',0}D^{J_2*}_{d,d'}D^{J_1}_{d,M}\nonumber\\
	&\quad+(-1)^{J_1+J_2}D^{J_3*}_{d',0}D^{J_2*}_{-d,d'}D^{J_1}_{-d,M}\text{.}
\end{align}
The coefficients $\gamma_{J_3}$, $\alpha^{J_1J_2}_{d}$ and $\epsilon^{J_3J_2}_{d'}$, which are independent of the angles in (\ref{eq:W2}), are defined as follows:
\begin{align}
\label{eq:c}
\gamma_{J_3}&=-\sqrt{3}\sum^{0,1}_{\alpha}|C_{\alpha}|^2(-1)^\alpha
	\langle1,1;\alpha,-\alpha|J_30\rangle\\
\alpha^{J_1J_2}_d&=\sqrt{35}(1-\frac{\delta_{d0}}{2})\nonumber\\
	&\quad\times\sum_{s(d)}\left[A_{\frac{s+d}{2}}A^*_{\frac{s-d}{2}}+(-1)^{J_2+J_1}
	A^*_{\frac{s+d}{2}}A_{\frac{s-d}{2}}\right]\nonumber\\
	&\quad\times\left\langle 3 3;\frac{s+d-2}{2},-\frac{s-d-2}{2}|J_1d\right\rangle\nonumber\\
	&\quad\times\left\langle2 2;\frac{s+d}{2},-\frac{s-d}{2}|J_2 d\right\rangle\\
s(d)&=-(4-|d|),-(4-|d|)+2,\text{...},(4-|d|)\nonumber\\
\label{eq:e}
  \epsilon^{J_3J_2}_{d'}&=\sqrt{15}(1-\frac{\delta_{d'0}}{2})\nonumber\\
  &\quad\times\sum_{s'(d')}\left[E_{\frac{s'+d'}{2}}E^*_{\frac{s'-d'}{2}}+(-1)^{J_2}E^*_{\frac{s'+d'}{2}}E_{\frac{s'-d'}{2}}\right]\nonumber\\
  &\quad\times\left\langle2 2 ;\frac{s'+d'}{2},-\frac{s'-d'}{2}|J_2 d'\right\rangle\nonumber\\
  &\quad\times\left\langle1 1;\frac{s'+d'-2}{2},-\frac{s'-d'-2}{2}|J_3 d'\right\rangle\\
s'(d')&=|d'|,|d'|+2,\text{...},4-|d'|\text{.}\nonumber
\end{align}
In (\ref{eq:W2}) the components of the polarization vectors are contained in the coefficients defined as follows:
\begin{align}
\label{eq:b}
\beta^{J_1}_0&=-\sqrt{7}\langle3 3;0 0|J_1 0\rangle|B_0|^2(P_-+P_A)\nonumber\\
	&\quad+\frac{\sqrt{7}}{2}\langle3 3;-1 1|J_1 0\rangle|B_1|^2
	[(P_+-P_{1z'}-P_{2z'})\nonumber\\
	&\qquad+(-1)^{J_1}(P_++P_{1z'}+P_{2z'})]\nonumber\\
\beta^{J_1}_1&=\frac{\sqrt{7}}{2}\langle 3 3;0 1|J_1 1\rangle\{\text{Re}(B_0B_1^*)
	[(P_{1x'}+P_{2x'}-P_E)\nonumber\\
		&\qquad-(-1)^{J_1}(P_{1x'}+P_{2x'}+P_E)]\nonumber\\
	&\quad+\text{Im}(B_0B_1^*)[(P_{1y'}+P_{2y'}-P_D)\nonumber\\
		&\qquad+(-1)^{J_1}(P_{1y'}+P_{2y'}+P_D)]\}\nonumber\\
	&\quad-\frac{\sqrt{7}}{2}\operatorname{i}\langle 3 3;0 1|J_11\rangle\{\text{Re}(B_0B_1^*)\nonumber\\
	&\quad\times[(P_{1y'}+P_{2y'}-P_D)\nonumber\\
		&\qquad-(-1)^{J_1}(P_{1y'}+P_{2y'}+P_D)]\nonumber\\
	&\quad-\text{Im}(B_0B_1^*)[(P_{1x'}+P_{2x'}-P_E)\nonumber\\
	&\qquad+(-1)^{J_1}(P_{1x'}+P_{2x'}+P_E)]\}\nonumber\\
\beta^{J_1}_2&=\frac{\sqrt{7}}{4}\langle33;1,1|J_12\rangle|B_1|^2\nonumber\\
	&\quad\times[1+(-1)^{J_1}](P_B-\operatorname{i}P_C)
 \end{align}
where
\begin{align}
	P_{\pm}&=1\pm P_{1z'}P_{2z'}\text{,}\nonumber\\
	P_A&=P_{1x'}P_{2x'}+P_{1y'}P_{2y'}\text{,}\nonumber\\
	P_B&=P_{1x'}P_{2x'}-P_{1y'}P_{2y'}\text{,}\nonumber\\
	P_C&=P_{1x'}P_{2y'}+P_{1y'}P_{2x'}\text{,}\nonumber\\
	P_D&=P_{1y'}P_{2z'}+P_{1z'}P_{2y'}\text{,}\nonumber\\
	P_E&=P_{1z'}P_{2x'}+P_{1x'}P_{2z'}\text{.}
\end{align}

The explicit expressions for the non-zero coefficients, $\gamma_{J_3}$, $\alpha^{J_1J_2}_{d}$, $\epsilon^{J_3J_2}_{d'}$ and $\beta^{J_1}_M$, in \eqref{eq:W2} are given in \ref{appendix}. Since the combined angular distribution in \eqref{eq:W2} is expressed as a sum of products of the orthogonal Wigner $D$-functions, we can obtain the values for these coefficients from
\begin{align}\label{eq:obtain coef}
&\gamma_{J_3}\alpha^{J_1J_2}_{d}\epsilon^{J_3J_2}_{d'}[\beta^{J_1}_M+(-1)^{J_1}\beta^{J_1*}_M]
	[1+(-1)^{J_1+J_2}\delta_{d0}]\nonumber\\
	&\quad\times[1+(-1)^{J_1+J_2}\delta_{d'0}\delta_{M0}]\nonumber\\
	&=(2J_1+1)(2J_2+1)(2J_3+1)(2-\delta_{M0})\nonumber\\
	&\quad\times\int W(\theta,\phi;\theta',\phi';\theta'',\phi'')\nonumber\\
	&\quad\times[Y^{J_1J_2J_3}_{dd'M}+Y^{J_1J_2J_3*}_{dd'M}]
	d\Omega d\Omega'd\Omega''\textbf{.}
\end{align}
In calculating \eqref{eq:obtain coef}, we made use of the orthogonality relation:
\begin{align}\label{eq:orthogonality relation}
&\int^{2\pi}_{0}d\alpha\int^{2\pi}_{0}d\gamma\int^{\pi}_{0}D^{j*}_{mm'}(\alpha,\beta,\gamma)
	D^{j'}_{\mu\mu'}(\alpha,\beta,\gamma)\sin\beta d\beta\nonumber\\
	&=\frac{8\pi^2}{(2j+1)}\delta_{m\mu}\delta_{m'\mu'}\delta_{jj'}\textbf{.}
\end{align}

When we have sufficient experimental data for the angular distribution function $W(\theta,\phi;\theta',\phi';\theta'',\phi'')$, the integral on the right side of  \eqref{eq:obtain coef} can be determined numerically for all possible allowed values of $J_1$, $J_2$, $J_3$, $d$, $d'$ and $M$, and hence the coefficients $\gamma_{J_3}$, $\alpha^{J_1J_2}_d$, $\epsilon^{J_3J_2}_{d'}$ and $\beta^{J_1}_M$ on the left side of \eqref{eq:obtain coef} can be obtained. Using the explicit expressions for these coefficients, this will give us $18$ independent equations to solve for the relative magnitudes as well as the relative phases of all the angular-momentum helicity amplitudes $A_i$ and $E_j$ in the radiative decay processes  ${}^3D_3\rightarrow{}^3P_2+\gamma_1$ and ${}^3P_2\rightarrow\psi+\gamma_2$, respectively, when either the initial proton or antiproton is polarized. Moreover, we can also obtain the relative magnitude and the relative phase of the two independent helicity amplitudes $B_0$ and $B_1$ in the initial process $\bar{p}p\rightarrow{}^3D_3$. It should be noted that the coefficients $\beta^{J_1}_M$ are functions of the longitudinal ($P_{z'}$) and the transverse ($P_{x'}$,$P_{y'}$) components of the polarization vectors of $\bar{p}$ and $p$. If the polarization vectors $\vec{P_1}$ and $\vec{P_2}$ go to zero, then $\beta^{L_1}_M=0$ when $M$ is nonzero or when $J_1$ is odd, and we will recover the results of the unpolarized $\bar{p}p$ collisions given in \cite{Our paper}.

\section{Partially integrated angular distributions}
The partially integrated angular distributions obtained from \eqref{eq:W2} will look a lot simpler and we will gain greater insight from them. We calculate six different cases of partially integrated angular distributions. In deriving these results, we frequently make use of \eqref{eq:orthogonality relation} and the following property of the $D$-functions:

\begin{flalign}
&\int_0^{2\pi}d\phi\int_0^{\pi}D^{J}_{MM'}(\phi,\theta,-\phi)\sin\theta d\theta\nonumber\\
&=\int_0^{2\pi}d\phi\int_0^{\pi}D^{J*}_{MM'}(\phi,\theta,-\phi)\sin\theta d\theta\nonumber\\
&=2\pi\delta_{M-M',0}\int_0^{\pi}d^J_{MM'}(\theta)
\sin\theta d\theta\nonumber\\
&=2\pi K_{JM}\text{,}
\end{flalign}
where
\begin{align}
K_{JM}=\int^{\pi}_0d^J_{MM}(\theta)\sin\theta d\theta\text{.}
\end{align}
We will express the final results for the three cases of single-particle angular distributions in terms of the orthogonal spherical harmonics by making use of the relation:
\begin{align}
D^{J}_{M0}=\sqrt{\frac{4\pi}{2J+1}}Y^*_{JM}\textbf{.}
\end{align}

Case $1$: We will integrate over ($\theta'$,$\phi'$) and ($\theta''$,$\phi''$). Only the angular distribution of the first gamma photon $\gamma_1$ is measured. We obtain
\begin{align}
\label{eq:Case1}
&\widetilde{W}(\theta,\phi)\nonumber\\
&=\int W(\theta,\phi,\theta',\phi',\theta'',\phi'')d\Omega'd\Omega'' \nonumber\\
&=\frac{1}{2\sqrt{\pi}}\gamma_0\epsilon^{00}_0\sum^{0,2,4,6}_{J_1}\sum^{\operatorname{min}\{J_1,2\}}_{M=0}(2-\delta_{M0})
	\alpha^{J_10}_0(-1)^M\nonumber\\
	&\quad\times\frac{1}{\sqrt{2J_1+1}}\operatorname{Re}(\beta^{J_1*}_M Y_{J_1M}(\theta,\phi))\nonumber\\
&=\frac{1}{2\sqrt{\pi}}\Bigg\{
	\frac{1}{2\sqrt{\pi}}[|B_0|^2(P_-+P_A)+|B_1|^2P_+]\nonumber\\
	&\quad-\frac{\sqrt{5}}{3}Y_{20}(\theta,\phi)
		\left[|B_0|^2(P_0+P_A)+\frac{3}{4}|B_1|^2P_+\right]\nonumber\\
		&\quad\times\left(|A_0|^2-\frac{3}{5}|A_2|^2-\frac{4}{5}|A_3|^2-\frac{3}{5}|A_4|^2\right)\nonumber\\
	&\quad-\sqrt{\frac{5}{3}}\operatorname{Re}(\beta^{2*}_1Y_{21}(\theta,\phi))\nonumber\\
		&\quad\times\left(|A_0|^2-\frac{3}{5}|A_2|^2-\frac{4}{5}|A_3|^2-\frac{3}{5}|A_4|^2\right)\nonumber\\
	&\quad+\sqrt{\frac{5}{3}}\operatorname{Re}(\beta^{2*}_2Y_{22}(\theta,\phi))\nonumber\\
		&\quad\times\left(|A_0|^2-\frac{3}{5}|A_2|^2-\frac{4}{5}|A_3|^2-\frac{3}{5}|A_4|^2\right)\nonumber\\
	&\quad+\frac{3}{11}Y_{40}(\theta,\phi)
		\left[|B_0|^2(P_-+P_A)+\frac{1}{6}|B_1|^2P_+\right]\nonumber\\
		&\quad\times\left(|A_0|^2-\frac{7}{3}|A_1|^2+\frac{1}{3}|A_2|^2+2|A_3|^2+\frac{1}{3}|A_4|^2\right)\nonumber\\
	&\quad-\sqrt{\frac{2}{11}}\operatorname{Re}(\beta^{4*}_1Y_{41}(\theta,\phi))\nonumber\\
		&\quad\times\left(|A_0|^2-\frac{7}{3}|A_1|^2+\frac{1}{3}|A_2|^2+2|A_3|^2+\frac{1}{3}|A_4|^2\right)\nonumber\\
	&\quad+\sqrt{\frac{2}{11}}\operatorname{Re}(\beta^{4*}_2Y_{42}(\theta,\phi))\nonumber\\
		&\quad\times\left(|A_0|^2-\frac{7}{3}|A_1|^2+\frac{1}{3}|A_2|^2+2|A_3|^2+\frac{1}{3}|A_4|^2\right)\nonumber\\
	&\quad-\frac{5}{33\sqrt{13}}Y_{60}(\theta,\phi)\left[|B_0|^2(P_-+P_A)-\frac{3}{4}|B_1|^2P_+\right]\nonumber\\
		&\quad\times\left(|A_0|^2-6|A_1|^2+15|A_2|^2-20|A_3|^2+15|A_4|^2\right)\nonumber\\
	&\quad-\frac{1}{\sqrt{429}}\operatorname{Re}(\beta^{6*}_1Y_{61}(\theta,\phi))\nonumber\\
		&\quad\times\left(|A_0|^2-6|A_1|^2+15|A_2|^2-20|A_3|^2+15|A_4|^2\right)\nonumber\\
	&\quad+\frac{1}{\sqrt{429}}\operatorname{Re}(\beta^{6*}_2Y_{62}(\theta,\phi))\nonumber\\
		&\quad\times\left(|A_0|^2-6|A_1|^2+15|A_2|^2-20|A_3|^2+15|A_4|^2\right)\Bigg\}\text{,}
\end{align}
where the angles ($\theta$,$\phi$) represent the direction of $\bar{p}$ measured from the $z$ axis, which is taken to be the direction of the momentum of ${}^3P_2$. This angle is the same as that of $\gamma_1$ measured in the ${}^3D_3$ rest frame with the $z$ axis taken to be the direction of the proton. The $x$ and $y$ axes are arbitrary. With the normalization condition $|B_0|^2+|B_1|^2=1$, \eqref{eq:Case1} allows us to determine the relative magnitude and the relative phase of the two helicity amplitudes in the process $\bar{p}p\rightarrow{}^3D_3$. There are also three equations relating the relative magnitudes of the $A$ helicity amplitudes.

Case $2$: We will integrate over ($\theta$,$\phi$) and ($\theta''$,$\phi''$). Only the angular distribution of the second gamma photon $\gamma_2$ is measured. We get
\begin{align}
\label{eq:Case2}
&\widetilde{W}(\theta',\phi')\nonumber\\
&=\int W(\theta,\phi;\theta',\phi';\theta'',\phi'')d\Omega d\Omega''\nonumber\\
&=\frac{1}{8\sqrt{\pi}}\sum^6_{J_1=0}\sum^{0,2,4}_{J_2}\sum^{\operatorname{min}\{J_1,J_2,2\}}_{d=0}(2-\delta_{d0})\nonumber\\
	&\quad\times\frac{1}{\sqrt{2J_2+1}}K_{J_1d}\alpha^{J_1J_2}_d\epsilon^{0J_2}_0\nonumber\\
	&\quad\times\left[\beta^{J_1}_dY_{J_2d}(\theta',\phi')+(-1)^{J_1}\beta^{J_1*}_dY^*_{J_2d}(\theta',\phi')\right]\nonumber\\
&=\frac{1}{4\sqrt{\pi}}\Bigg\{\frac{1}{\sqrt{\pi}}[|B_0|^2(P_-+P_A)+|B_1|^2P_+]\nonumber\\
	&\quad-\sqrt{\frac{2}{7}}\left(|E_0|^2+\frac{1}{2}|E_1|^2-|E_2|^2\right)\bigg\{
		2\sqrt{\frac{10}{7}}Y_{20}(\theta',\phi')\nonumber\\
		&\quad\times[|B_0|^2(P_-+P_A)+|B_1|^2P_+]\nonumber\\
		&\quad\times\left(|A_0|^2-\frac{1}{2}|A_1|^2-|A_2|^2-\frac{1}{2}|A_3|^2+|A_4|^2\right)\nonumber\\
		&\quad-\frac{10}{3}\sqrt{\frac{5}{7}}\operatorname{Re}(\beta^2_1Y_{21}(\theta',\phi'))	
			\Big[\text{Re}(A_1A_0^*)\nonumber\\
			&\qquad+\frac{1}{\sqrt{10}}\text{Re}(A_2A_1^*)-\frac{1}{5\sqrt{3}}\text{Re}(A_3A_2^*)\nonumber\\
			&\qquad+\frac{\sqrt{2}}{5}\text{Re}(A_4A_3^*)\Big]\nonumber\\
		&\quad+\frac{40}{3\sqrt{21}}\operatorname{Re}(\beta^2_2Y_{22}(\theta',\phi')) \nonumber\\
			&\quad\times\Big[\text{Re}(A_2A_0^*)+\sqrt{3}\text{Re}(A_3A_1^*)+2\sqrt{\frac{3}{5}}\text{Re}(A_4A_2^*)\Big]\nonumber\\
		&\quad-\frac{6}{\sqrt{77}}\operatorname{Re}(\beta^4_1Y_{21}(\theta',\phi'))
			\Big[\text{Re}(A_1A_0^*)\nonumber\\
				&\qquad-\frac{2}{3}\sqrt{\frac{2}{5}}\text{Re}(A_2A_1^*)
				+\frac{1}{2\sqrt{3}}\text{Re}(A_3A_2^*)\nonumber\\
				&\qquad-\frac{1}{\sqrt{2}}\text{Re}(A_4A_3^*)\Big]\nonumber\\
		&\quad+\frac{12}{5}\sqrt{\frac{30}{77}}\operatorname{Re}(\beta^4_2Y_{22}(\theta',\phi'))
			\Big[\text{Re}(A_2A_0^*)\nonumber\\
				&\qquad-\frac{1}{2\sqrt{3}}\text{Re}(A_3A_1^*)-\frac{2}{3}\sqrt{\frac{5}{3}}\text{Re}(A_4A_2^*)\Big]\nonumber\\
		&\quad-\frac{2}{21}\sqrt{\frac{5}{11}}\operatorname{Re}(\beta^6_1Y_{21}(\theta',\phi'))
			\Big[\text{Re}(A_1A_0^*)\nonumber\\
				&\qquad-\sqrt{\frac{5}{2}}\text{Re}(A_2A_1^*)-\frac{5}{\sqrt{3}}\text{Re}(A_3A_2^*)\nonumber\\
				&\qquad+\frac{10}{\sqrt{2}}\text{Re}(A_4A_3^*)\Big]\nonumber\\
		&\quad+\frac{8}{21}\sqrt{\frac{10}{33}}\operatorname{Re}(\beta^6_2Y_{22}(\theta',\phi'))\nonumber\\
			&\quad\times\Big[\text{Re}(A_2A_0^*)-2\sqrt{3}\text{Re}(A_3A_1^*)+\sqrt{15}\text{Re}(A_4A_2^*)\Big]\nonumber\\
		&\quad+6\sqrt{\frac{5}{7}}\operatorname{Im}(\beta^1_1Y_{21}(\theta',\phi'))
			\Big[\text{Im}(A_1A_0^*)\nonumber\\
			&\qquad+\frac{1}{3}\sqrt{\frac{5}{2}}\text{Im}(A_2A_1^*)
				-\frac{1}{\sqrt{3}}\text{Im}(A_3A_2^*)\nonumber\\
				&\qquad-\sqrt{2}\text{Im}(A_4A_3^*)\Big]\nonumber\\
		&\quad+\frac{2\sqrt{5}}{3}\operatorname{Im}(\beta^3_1Y_{21}(\theta',\phi'))\nonumber\\
			&\quad\times\Big[\text{Im}(A_1A_0^*)+\frac{1}{2\sqrt{3}}\text{Im}(A_3A_2^*)+\frac{1}{\sqrt{2}}\text{Im}(A_4A_3^*)\Big]\nonumber\\
		&\quad+\frac{2}{3\sqrt{7}}\operatorname{Im}(\beta^5_1Y_{21}(\theta',\phi'))
			\Big[\text{Im}(A_1A_0^*)\nonumber\\
			&\qquad-\frac{3}{\sqrt{10}}\text{Im}(A_2A_1^*)-\frac{1}{\sqrt{3}}\text{Im}(A_3A_2^*)\nonumber\\
			&\qquad-\sqrt{2}\text{Im}(A_4A_3^*)\Big]\bigg\}\nonumber\\
	&\quad+\sqrt{\frac{2}{7}}\left(|E_0|^2-\frac{2}{3}|E_1|^2+\frac{1}{6}|E_2|^2\right)\bigg\{
		\sqrt{\frac{2}{7}}Y_{40}(\theta',\phi')\nonumber\\
			&\quad\times[|B_0|^2(P_-+P_A)+|B_1|^2P_+]\nonumber\\
			&\quad\times(|A_0|^2-4|A_1|^2+6|A_2|^2-4|A_3|^2+|A_4|^2)\nonumber\\
		&\quad-\frac{5}{3}\sqrt{\frac{10}{21}}\operatorname{Re}(\beta^2_1Y_{41}(\theta',\phi'))
			\Big[\text{Re}(A_1A_0^*)\nonumber\\
			&\qquad-3\sqrt{\frac{2}{5}}\text{Re}(A_2A_1^*)+\frac{2\sqrt{3}}{5}\text{Re}(A_3A_2^*)\nonumber\\
			&\qquad+\frac{\sqrt{2}}{5}\text{Re}(A_4A_3^*)\Big]\nonumber\\
		&\quad+\frac{20}{3\sqrt{7}}\operatorname{Re}(\beta^2_2Y_{42}(\theta',\phi'))\nonumber\\
			&\quad\times\Big[\text{Re}(A_2A_0^*)-\frac{4}{\sqrt{3}}\text{Re}(A_3A_1^*)+2\sqrt{\frac{3}{5}}\text{Re}(A_4A_2^*)\Big]\nonumber\\
		&\quad-\sqrt{\frac{6}{77}}\operatorname{Re}(\beta^4_1Y_{41}(\theta',\phi'))
			\Big[\text{Re}(A_1A_0^*)\nonumber\\
			&\qquad+4\sqrt{\frac{2}{5}}\text{Re}(A_2A_1^*)-\sqrt{3}\text{Re}(A_3A_2^*)\nonumber\\
			&\qquad-\frac{1}{\sqrt{2}}\text{Re}(A_4A_3^*)\Big]\nonumber\\
		&\quad+\frac{18}{5}\sqrt{\frac{10}{77}}\operatorname{Re}(\beta^4_2Y_{42}(\theta',\phi'))\nonumber\\
			&\quad\times\Big[\text{Re}(A_2A_0^*)+\frac{2}{3\sqrt{3}}\text{Re}(A_3A_1^*)
				-\frac{2}{3}\sqrt{\frac{5}{3}}\text{Re}(A_4A_2^*)\Big]\nonumber\\
		&\quad-\frac{1}{21}\sqrt{\frac{10}{33}}\operatorname{Re}(\beta^6_1Y_{41}(\theta',\phi'))
			\Big[\text{Re}(A_1A_0^*)\nonumber\\
			&\qquad+3\sqrt{10}\text{Re}(A_2A_1^*)+10\sqrt{3}\text{Re}(A_3A_2^*)\nonumber\\
			&\qquad+5\sqrt{2}\text{Re}(A_4A_3^*)\Big]\nonumber\\
		&\quad+\frac{4}{21}\sqrt{\frac{10}{11}}\operatorname{Re}(\beta^6_2Y_{42}(\theta',\phi'))\nonumber\\
			&\quad\times\Big[\text{Re}(A_2A_0^*)+\frac{8}{\sqrt{3}}\text{Re}(A_3A_1^*)+\sqrt{15}\text{Re}(A_4A_2^*)\Big]\nonumber\\
		&\quad+\sqrt{\frac{30}{7}}\operatorname{Im}(\beta^1_1Y_{41}(\theta',\phi'))
			\Big[\text{Im}(A_1A_0^*)\nonumber\\
			&\qquad-\sqrt{10}\text{Im}(A_2A_1^*)+2\sqrt{3}\text{Im}(A_3A_2^*)\nonumber\\
			&\qquad-\sqrt{2}\text{Im}(A_4A_3^*)\Big]\nonumber\\
		&\quad+\frac{1}{3}\sqrt{\frac{10}{3}}\operatorname{Im}(\beta^3_1Y_{41}(\theta',\phi'))\nonumber\\
			&\quad\times\Big[\text{Im}(A_1A_0^*)-\sqrt{3}\text{Im}(A_3A_2^*)+\frac{1}{\sqrt{2}}\text{Im}(A_4A_3^*)\Big]\nonumber\\
		&\quad+\frac{1}{3}\sqrt{\frac{2}{21}}\operatorname{Im}(\beta^5_1Y_{41}(\theta',\phi'))
			\Big[\text{Im}(A_1A_0^*)\nonumber\\
			&\qquad+9\sqrt{\frac{2}{5}}\text{Im}(A_2A_1^*)+2\sqrt{3}\text{Im}(A_3A_2^*)\nonumber\\
			&\qquad-\sqrt{2}\text{Im}(A_4A_3^*)\Big]\bigg\}\Bigg\}\text{.}
\end{align}
Here, ($\theta'$,$\phi'$) are the angles between ${}^3D_3$ and $\gamma_2$ in the ${}^3P_2$ rest frame. As we can obtain one equation relating the relative magnitudes of the $E$ helicity amplitudes from case $3$ and also three equations relating the relative magnitudes of the $A$ helicity amplitudes from case $1$, the measurement of the single-particle angular distribution of $\gamma_2$ allows us to determine the relative magnitudes of the $E$ and $A$ helicity amplitudes and also the cosines of the relative phases of the $A$ helicity amplitudes. It should be noted that $\beta^{J_1}_M$($M\neq0$) will vanish if there is no polarization in the $p$ and $\bar{p}$ beams, and we will not get any information on the helicity amplitudes \cite{Our paper}. So the polarization of the proton or the antiproton is crucial for extracting this information from the single-particle angular distributions.

Case $3$: We will integrate over ($\theta$,$\phi$) and ($\theta'$,$\phi'$). Only the angular distribution of the electron is measured. We have
\begin{align}
\label{eq:Case3}
&\widetilde{W}(\theta'',\phi'')\nonumber\\
&=\int W(\theta,\phi,\theta',\phi',\theta'',\phi'')d\Omega d\Omega'\nonumber\\
&=\frac{1}{64\sqrt{\pi}}\sum^6_{J_1=0}\sum^4_{J_2=0}\sum^{0,2}_{J_3}
	\sum^{\operatorname{min}\{J_1,J_2,J_3\}}_{d=0}(1+\delta_{d0})^2\frac{1}{\sqrt{2J_3+1}}\nonumber\\
	&\quad\times K_{J_1d}K_{J_2d}\gamma_{J_3}\alpha^{J_1J_2}_d\epsilon^{J_3J_2}_d\nonumber\\
	&\quad\times\left[\beta^{J_1}_dY_{J_3d}(\theta'',\phi'')+(-1)^{J_1}\beta^{J_1*}_dY^*_{J_3d}(\theta'',\phi'')\right]\nonumber\\
&=\frac{1}{32\sqrt{\pi}}\Bigg\{\frac{8}{\sqrt{\pi}}[|B_0|^2(P_-+P_A)+|B_1|^2|P_+]\nonumber\\
&\quad+\frac{8}{\sqrt{5}}Y_{20}(\theta'',\phi'')[|B_0|^2(P_-+P_A)+|B_1|^2|P_+]\nonumber\\
	&\quad\times(|E_0|^2-2|E_1|^2+|E_2|^2)\nonumber\\
&\quad+\operatorname{Re}(\beta^2_1Y_{21}(\theta'',\phi''))\bigg\{\sqrt{\frac{5}{6}}
	\Big[\text{Im}(E_1E_0^*)\nonumber\\
	&\qquad-\sqrt{\frac{2}{3}}\text{Im}(E_2E_1^*)\Big]
	\Big[\text{Im}(A_1A_0^*)+\frac{3}{\sqrt{10}}\text{Im}(A_2A_1^*)\nonumber\\
		&\qquad+\frac{\sqrt{3}}{5}\text{Im}(A_3A_2^*)-\frac{\sqrt{2}}{5}\text{Im}(A_4A_3^*)\Big]\nonumber\\
	&\qquad-\frac{5}{21}\sqrt{\frac{5}{6}}\Big[\text{Re}(E_1E_0^*)-\sqrt{6}\text{Re}(E_2E_1^*)\Big]
		\Big[\text{Re}(A_1A_0^*)\nonumber\\
		&\qquad+\frac{1}{\sqrt{10}}\text{Re}(A_2A_1^*)-\frac{1}{5\sqrt{3}}\text{Re}(A_3A_2^*)\nonumber\\
		&\qquad+\frac{\sqrt{2}}{5}\text{Re}(A_4A_3^*)\Big]\nonumber\\
	&\qquad-\frac{1}{6}\sqrt{\frac{5}{6}}\Big[\text{Im}(E_1E_0^*)+\sqrt{\frac{3}{2}}\text{Im}(E_2E_1^*)\Big]
		\Big[\text{Im}(A_1A_0^*)\nonumber\\
		&\qquad-\sqrt{\frac{2}{5}}\text{Im}(A_2A_1^*)-\frac{2}{5\sqrt{3}}\text{Im}(A_3A_2^*)\nonumber\\
		&\qquad-\frac{\sqrt{2}}{5}\text{Im}(A_4A_3^*)\Big]\nonumber\\
	&\qquad+\frac{1}{14}\sqrt{\frac{5}{6}}\Big[\text{Re}(E_1E_0^*)+\sqrt{\frac{1}{6}}\text{Re}(E_2E_1^*)\Big]
		\Big[\text{Re}(A_1A_0^*)\nonumber\\
		&\qquad-3\sqrt{\frac{2}{5}}\text{Re}(A_2A_1^*)
			+\frac{2\sqrt{3}}{5}\text{Re}(A_3A_2^*)\nonumber\\
		&\qquad+\frac{\sqrt{2}}{5}\text{Re}(A_4A_3^*)\Big]\bigg\}\nonumber\\
&\quad+\operatorname{Re}(\beta^2_2Y_{22}(\theta'',\phi''))\bigg\{\frac{80}{63}\text{Re}(E_2E_0^*)\nonumber\\
	&\qquad\times\Big[\text{Re}(A_2A_0^*)+\sqrt{3}\text{Re}(A_3A_1^*)+2\sqrt{\frac{3}{5}}\text{Re}(A_4A_2^*)\Big]\nonumber\\
	&\qquad-\frac{10}{3}\sqrt{\frac{10}{3}}\text{Im}(E_2E_0^*)\Big[\text{Im}(A_2A_0^*)\nonumber\\
	&\qquad-2\sqrt{\frac{3}{5}}\text{Im}(A_4A_2^*)\Big]\nonumber\\
	&\qquad+\frac{2}{7}\text{Re}(E_2E_0^*)\Big[\text{Re}(A_2A_0^*)-\frac{4}{\sqrt{3}}\text{Re}(A_3A_1^*)\nonumber\\
		&\qquad+2\sqrt{\frac{3}{5}}\text{Re}(A_4A_2^*)\Big]\bigg\}\nonumber\\
&\quad+\operatorname{Re}(\beta^4_1Y_{21}(\theta'',\phi''))\bigg\{\frac{3}{10}\sqrt{\frac{6}{11}}
		\Big[\text{Im}(E_1E_0^*)\nonumber\\
		&\qquad-\sqrt{\frac{2}{3}}\text{Im}(E_2E_1^*)\Big]
		\Big[\text{Im}(A_1A_0^*)-2\sqrt{\frac{2}{5}}\text{Im}(A_2A_1^*)\nonumber\\
		&\qquad-\frac{\sqrt{3}}{2}\text{Im}(A_3A_2^*)
			+\frac{1}{\sqrt{2}}\text{Im}(A_4A_3^*)\Big]\nonumber\\
	&\qquad-\frac{1}{7}\sqrt{\frac{3}{22}}\Big[\text{Re}(E_1E_0^*)-\sqrt{6}\text{Re}(E_2E_1^*)\Big]
		\Big[\text{Re}(A_1A_0^*)\nonumber\\
		&\qquad-\frac{2}{3}\sqrt{\frac{2}{5}}\text{Re}(A_2A_1^*)+\frac{1}{2\sqrt{3}}\text{Re}(A_3A_2^*)\nonumber\\
		&\qquad-\frac{1}{\sqrt{2}}\text{Re}(A_4A_3^*)\Big]\nonumber\\
	&\qquad-\frac{1}{10}\sqrt{\frac{3}{22}}\Big[\text{Im}(E_1E_0^*)+\sqrt{\frac{3}{2}}\text{Im}(E_2E_1^*)\Big]\nonumber\\
		&\qquad\times\Big[\text{Im}(A_1A_0^*)+\frac{4}{3}\sqrt{\frac{2}{5}}\text{Im}(A_2A_1^*)
			+\frac{1}{\sqrt{3}}\text{Im}(A_3A_2^*)\nonumber\\
			&\qquad+\frac{1}{\sqrt{2}}\text{Im}(A_4A_3^*)\Big]\nonumber\\
	&\qquad+\frac{3}{140}\sqrt{\frac{6}{11}}\Big[\text{Re}(E_1E_0^*)+\sqrt{\frac{1}{6}}\text{Re}(E_2E_1^*)\Big]\nonumber\\
		&\qquad\times\Big[\text{Re}(A_1A_0^*)+4\sqrt{\frac{2}{5}}\text{Re}(A_2A_1^*)-\sqrt{3}\text{Re}(A_3A_2^*)\nonumber\\
			&\qquad-\frac{1}{\sqrt{2}}\text{Re}(A_4A_3^*)\Big]\bigg\}\nonumber\\
&\quad+\operatorname{Re}(\beta^4_2Y_{22}(\theta'',\phi''))\bigg\{\frac{24}{7}\sqrt{\frac{2}{55}}\text{Re}(E_2E_0^*)
		\Big[\text{Re}(A_2A_0^*)\nonumber\\
		&\qquad-\frac{1}{2\sqrt{3}}\text{Re}(A_3A_1^*)-\frac{2}{3}\sqrt{\frac{5}{3}}\text{Re}(A_4A_2^*)\Big]\nonumber\\
	&\qquad-3\sqrt{\frac{2}{55}}\text{Im}(E_2E_0^*)
		\Big[\text{Im}(A_2A_0^*)+\frac{2}{3}\sqrt{\frac{5}{3}}\text{Im}(A_4A_2^*)\Big]\nonumber\\
	&\qquad+\frac{27}{35}\sqrt{\frac{2}{55}}\text{Re}(E_2E_0^*)
		\Big[\text{Re}(A_2A_0^*)+\frac{2}{3\sqrt{3}}\text{Re}(A_3A_1^*)\nonumber\\
		&\qquad-\frac{2}{3}\sqrt{\frac{5}{3}}\text{Re}(A_4A_2^*)\Big]\bigg\}\nonumber\\
&\quad+\operatorname{Re}(\beta^6_1Y_{21}(\theta'',\phi''))\bigg\{\frac{1}{\sqrt{2310}}\Big[\text{Im}(E_1E_0^*)\nonumber\\
	&\qquad-\sqrt{\frac{2}{3}}\text{Im}(E_2E_1^*)\Big]
		\Big[\text{Im}(A_1A_0^*)-3\sqrt{\frac{5}{2}}\text{Im}(A_2A_1^*)\nonumber\\
		&\qquad+5\sqrt{3}\text{Im}(A_3A_2^*)-\frac{10}{\sqrt{2}}\text{Im}(A_4A_3^*)\Big]\nonumber\\
	&\qquad-\frac{1}{21}\sqrt{\frac{5}{462}}\Big[\text{Re}(E_1E_0^*)-\sqrt{6}\text{Re}(E_2E_1^*)\Big]\nonumber\\
		&\qquad\times\Big[\text{Re}(A_1A_0^*)-\sqrt{\frac{5}{2}}\text{Re}(A_2A_1^*)-\frac{5}{\sqrt{3}}\text{Re}(A_3A_2^*)\nonumber\\
		&\qquad+\frac{10}{\sqrt{2}}\text{Re}(A_4A_3^*)\Big]\nonumber\\
	&\qquad-\frac{1}{42}\sqrt{\frac{7}{330}}\Big[\text{Im}(E_1E_0^*)+\sqrt{\frac{3}{2}}\text{Im}(E_2E_1^*)\Big]\nonumber\\
		&\qquad\times\Big[\text{Im}(A_1A_0^*)+\sqrt{10}\text{Im}(A_2A_1^*)-\frac{10}{\sqrt{3}}\text{Im}(A_3A_2^*)\nonumber\\
		&\qquad-5\sqrt{2}\text{Im}(A_4A_3^*)\Big]\nonumber\\
	&\qquad+\frac{1}{210}\sqrt{\frac{15}{154}}\Big[\text{Re}(E_1E_0^*)+\sqrt{\frac{1}{6}}\text{Re}(E_2E_1^*)\Big]\nonumber\\
		&\qquad\times\Big[\text{Re}(A_1A_0^*)+3\sqrt{10}\text{Re}(A_2A_1^*)+10\sqrt{3}\text{Re}(A_3A_2^*)\nonumber\\
			&\qquad+5\sqrt{2}\text{Re}(A_4A_3^*)\Big]\bigg\}\nonumber\\
&\quad+\operatorname{Re}(\beta^6_2Y_{22}(\theta'',\phi''))\bigg\{\frac{16}{63}\sqrt{\frac{10}{77}}\text{Re}(E_2E_0^*)
	\Big[\text{Re}(A_2A_0^*)\nonumber\\
		&\qquad-2\sqrt{3}\text{Re}(A_3A_1^*)+\sqrt{15}\text{Re}(A_4A_2^*)\Big]\nonumber\\
	&\qquad-\frac{2}{9}\sqrt{\frac{10}{77}}\text{Im}(E_2E_0^*)
		\Big[\text{Im}(A_2A_0^*)-\sqrt{15}\text{Im}(A_4A_2^*)\Big]\nonumber\\
	&\qquad+\frac{2}{35}\sqrt{\frac{10}{77}}\text{Re}(E_2E_0^*)
		\Big[\text{Re}(A_2A_0^*)+\frac{8}{\sqrt{3}}\text{Re}(A_3A_1^*)\nonumber\\
		&\qquad+\sqrt{15}\text{Re}(A_4A_2^*)\Big]\bigg\}\nonumber\\
&\quad+\operatorname{Im}(\beta^1_1Y_{21}(\theta'',\phi''))\bigg\{3\sqrt{\frac{3}{10}}
	\Big[\text{Im}(E_1E_0^*)\nonumber\\
		&\qquad-\sqrt{\frac{2}{3}}\text{Im}(E_2E_1^*)\Big]
		\Big[\text{Re}(A_1A_0^*)+\sqrt{\frac{5}{2}}\text{Re}(A_2A_1^*)\nonumber\\
			&\qquad+\sqrt{3}\text{Re}(A_3A_2^*)+\sqrt{2}\text{Re}(A_4A_3^*)\Big]\nonumber\\
	&\qquad+\frac{1}{7}\sqrt{\frac{15}{2}}\Big[\text{Re}(E_1E_0^*)-\sqrt{6}\text{Re}(E_2E_1^*)\Big]
		\Big[\text{Im}(A_1A_0^*)\nonumber\\
		&\qquad+\frac{1}{3}\sqrt{\frac{5}{2}}\text{Im}(A_2A_1^*)-\frac{1}{\sqrt{3}}\text{Im}(A_3A_2^*)\nonumber\\
		&\qquad-\sqrt{2}\text{Im}(A_4A_3^*)\Big]\nonumber\\
	&\qquad-\frac{1}{2}\sqrt{\frac{3}{10}}\Big[\text{Im}(E_1E_0^*)+\sqrt{\frac{3}{2}}\text{Im}(E_2E_1^*)\Big]
		\Big[\text{Re}(A_1A_0^*)\nonumber\\
		&\qquad-\frac{\sqrt{10}}{3}\text{Re}(A_2A_1^*)-\frac{2}{\sqrt{3}}\text{Re}(A_3A_2^*)\nonumber\\
		&\qquad+\sqrt{2}\text{Re}(A_4A_3^*)\Big]\nonumber\\
	&\qquad-\frac{3}{70}\sqrt{\frac{15}{2}}\Big[\text{Re}(E_1E_0^*)+\sqrt{\frac{1}{6}}\text{Re}(E_2E_1^*)\Big]\nonumber\\
		&\qquad\times\Big[\text{Im}(A_1A_0^*)-\sqrt{10}\text{Im}(A_2A_1^*)+2\sqrt{3}\text{Im}(A_3A_2^*)\nonumber\\
		&\qquad-\sqrt{2}\text{Im}(A_4A_3^*)\Big]\bigg\}\nonumber\\
&\quad+\operatorname{Im}(\beta^3_1Y_{21}(\theta'',\phi''))\bigg\{\sqrt{\frac{7}{30}}
	\Big[\text{Im}(E_1E_0^*)\nonumber\\
	&\qquad-\sqrt{\frac{2}{3}}\text{Im}(E_2E_1^*)\Big]
		\Big[\text{Re}(A_1A_0^*)-\frac{\sqrt{3}}{2}\text{Re}(A_3A_2^*)\nonumber\\
		&\qquad-\sqrt{2}\text{Re}(A_4A_3^*)\Big]\nonumber\\		
	&\qquad+\frac{1}{3}\sqrt{\frac{5}{42}}\Big[\text{Re}(E_1E_0^*)-\sqrt{6}\text{Re}(E_2E_1^*)\Big]\nonumber\\
		&\qquad\times\Big[\text{Im}(A_1A_0^*)+\frac{1}{2\sqrt{3}}\text{Im}(A_3A_2^*)+\frac{1}{\sqrt{2}}\text{Im}(A_4A_3^*)\Big]\nonumber\\
	&\qquad-\frac{1}{6}\sqrt{\frac{7}{30}}\Big[\text{Im}(E_1E_0^*)+\sqrt{\frac{3}{2}}\text{Im}(E_2E_1^*)\Big]\nonumber\\
		&\qquad\times\Big[\text{Re}(A_1A_0^*)+\frac{1}{\sqrt{3}}\text{Re}(A_3A_2^*)-\frac{1}{\sqrt{2}}\text{Re}(A_4A_3^*)\Big]\nonumber\\
	&\qquad-\frac{1}{2\sqrt{210}}\Big[\text{Re}(E_1E_0^*)+\sqrt{\frac{1}{6}}\text{Re}(E_2E_1^*)\Big]\nonumber\\
		&\qquad\times\Big[\text{Im}(A_1A_0^*)-\sqrt{3}\text{Im}(A_3A_2^*)+\frac{1}{\sqrt{2}}\text{Im}(A_4A_3^*)\Big]\bigg\}\nonumber\\
&\quad+\operatorname{Im}(\beta^5_1Y_{21}(\theta'',\phi''))\bigg\{\frac{1}{15}\sqrt{\frac{3}{2}}
		\Big[\text{Im}(E_1E_0^*)\nonumber\\
			&\qquad-\sqrt{\frac{2}{3}}\text{Im}(E_2E_1^*)\Big]
		\Big[\text{Re}(A_1A_0^*)-\frac{9}{\sqrt{10}}\text{Re}(A_2A_1^*)\nonumber\\
			&\qquad+\sqrt{3}\text{Re}(A_3A_2^*)+\sqrt{2}\text{Re}(A_4A_3^*)\Big]\nonumber\\
	&\qquad+\frac{1}{21\sqrt{6}}\Big[\text{Re}(E_1E_0^*)-\sqrt{6}\text{Re}(E_2E_1^*)\Big]
		\Big[\text{Im}(A_1A_0^*)\nonumber\\
			&\qquad-\frac{3}{\sqrt{10}}\text{Im}(A_2A_1^*)-\frac{1}{\sqrt{3}}\text{Im}(A_3A_2^*)\nonumber\\
			&\qquad-\sqrt{2}\text{Im}(A_4A_3^*)\Big]\nonumber\\
	&\qquad-\frac{1}{30\sqrt{6}}\Big[\text{Im}(E_1E_0^*)+\sqrt{\frac{3}{2}}\text{Im}(E_2E_1^*)\Big]
		\Big[\text{Re}(A_1A_0^*)\nonumber\\
		&\qquad+3\sqrt{\frac{2}{5}}\text{Re}(A_2A_1^*)-\frac{2}{\sqrt{3}}\text{Re}(A_3A_2^*)\nonumber\\
		&\qquad+\sqrt{2}\text{Re}(A_4A_3^*)\Big]\nonumber\\
	&\qquad-\frac{1}{70\sqrt{6}}\Big[\text{Re}(E_1E_0^*)+\sqrt{\frac{1}{6}}\text{Re}(E_2E_1^*)\Big]
		\Big[\text{Im}(A_1A_0^*)\nonumber\\
		&\qquad+9\sqrt{\frac{2}{5}}\text{Im}(A_2A_1^*)+2\sqrt{3}\text{Im}(A_3A_2^*)\nonumber\\
		&\qquad-\sqrt{2}\text{Im}(A_4A_3^*)\Big]\bigg\}\Bigg\}\text{,}
\end{align}
where ($\theta''$,$\phi''$) are the angles between the directions of the momenta of $e^-$ and ${}^3P_2$ in the $\psi$ rest frame. From the measurement of the angular distribution of the electron alone we find that we cannot get any more useful information on the helicity amplitudes.

So from cases $1$--$3$, we see that we can obtain the relative magnitudes of all the helicity amplitudes in the processes ${}^3D_3\rightarrow{}^3P_2+\gamma_1$ and ${}^3P_2\rightarrow\psi+\gamma_2$ by measuring only the single-particle angular distributions of $\gamma_1$, $\gamma_2$ and $e^-$. We can also get the cosines of the relative phases of the helicity amplitudes in the process ${}^3D_3\rightarrow{}^3P_2+\gamma_1$. In order to obtain complete information on the relative phases of the helicity amplitudes in the radiative processes, we need to measure the simultaneous angular distributions of two particles.

Case $4$: We will integrate over the angles ($\theta''$,$\phi''$), the direction of the final electron. The combined angular distribution of the two photons $\gamma_1$ and $\gamma_2$ is measured. We get
\begin{align}\label{eq:Case4}
&\widetilde{\widetilde{W}}(\theta,\phi,\theta',\phi')\nonumber\\
&=\int W(\theta,\phi,\theta',\phi',\theta'',\phi'')d\Omega''\nonumber\\
&=\frac{\gamma_0}{4(4\pi)^2}\sum^6_{J_1=0}\sum^{0,2,4}_{J_2}
	\sum^{\operatorname{min}\{J_1,J_2,2\}}_{d=0}
	\sum^{\operatorname{min}\{J_1,2\}}_{M=-\operatorname{min}\{J_1,2\}}(1+\delta_{M0})\nonumber\\
	&\quad\times\alpha^{J_1J_2}_{d}\epsilon^{0J_2}_0\left[\beta^{J_1}_MD^{J_1}_{dM}(\theta,\phi)D^{J_2*}_{d0}(\theta',\phi')\right.\nonumber\\
		&\quad\left.+(-1)^{J_1}\beta^{J_1*}_MD^{J_1*}_{dM}(\theta,\phi)D^{J_2}_{d0}(\theta',\phi')\right]
\end{align}

Since the explicit expressions for the partially inegrated angular distributions of two particles are rather long, we only give the results in terms of the sums of the coefficients defined in \ref{appendix}. In \eqref{eq:Case4}, however, we can obtain the coefficients of the angular functions from
\begin{align}\label{eq:coefficients}
&\alpha^{J_1J_2}_{d}\epsilon^{0J_2}_0[1+\delta_{M0}][1+\delta_{d0}\delta_{M0}]
	[\beta^{J_1}_{M}+(-1)^{J_1}\beta^{J_1*}_{M}]\nonumber\\
	&=4(2J_1+1)(2J_2+1)(2-\delta_{M0})\int\widetilde{\widetilde{W}}(\theta,\phi;\theta',\phi')\nonumber\\
		&\quad\times[D^{J_2*}_{d0}D^{J_1}_{dM}+D^{J_2}_{d0}D^{J_1*}_{dM}]d\Omega d\Omega'\text{.}
\end{align}
A close examination of the expressions for $\alpha^{L_1L_2}_d$ and $\epsilon^{0L_2}_0$ shows that \eqref{eq:coefficients} enables us to obtain the sines and the cosines of the relative phases of all the $A$ and $B$ helicity amplitudes. It also enables us to determine the relative magnitudes of all the $A$, $E$ and $B$ helicity amplitudes. Only the relative phases among the $E$ helicity amplitudes remain undetermined. We can get these phases only by measuring the simultaneous angular distribution of $\gamma_2$ and of $e^-$ as we will see in case $6$.

Case $5$: Here we integrate over ($\theta'$,$\phi'$) or the direction of $\gamma_2$ to get the combined angular distribution of $\gamma_1$ and $e^-$. We get
\begin{align}
&\widetilde{\widetilde{W}}(\theta,\phi;\theta'',\phi'')\nonumber\\
&=\int W(\theta,\phi;\theta',\phi';\theta'',\phi'') d\Omega'\nonumber\\
&=\frac{1}{16(4\pi)^2}\sum^6_{J_1=0}\sum^4_{J_2=0}\sum^{0,2}_{J_3}\sum^{\operatorname{min}\{J_1,J_2,J_3\}}_{d'}(1+\delta_{d'0})\nonumber\\
	&\quad\times\gamma_{J_3}\alpha^{J_1J_2}_{d'}\epsilon^{J_3J_2}_{d'}K_{J_2d'}
	\sum^{\operatorname{min}\{J_1,2\}}_{M=-\operatorname{min}\{J_1,2\}}(1+\delta_{M0})\nonumber\\
	&\quad\times[\beta^{J_1}_MD^{J_1}_{d'M}(\theta,\phi)D^{J_3*}_{d'0}(\theta'',\phi'')\nonumber\\
		&\quad+(-1)^{J_1}\beta^{J_1*}_MD^{J_1*}_{d'M}(\theta,\phi)D^{J_3}_{d'0}(\theta'',\phi'')]\text{.}
\end{align}
It turns out that we cannot obtain any useful information from this angular distribution.

Case $6$: This time we will integrate over the angles ($\theta$,$\phi$) to obtain the combined angular distribution of $\gamma_2$ and $e^-$ alone. We obtain
\begin{align}
\label{eq:Case6}
&\widetilde{\widetilde{W}}(\theta',\phi',\theta'',\phi'')\nonumber\\
&=\int W(\theta,\phi,\theta',\phi',\theta'',\phi'')d\Omega\nonumber\\
&=\frac{1}{4(4\pi)^2}\sum^6_{J_1=0}\sum^{4}_{J_2=0}\sum^{0,2}_{J_3}
	\sum^{\operatorname{min}\{J_1,J_2,2\}}_{d=-\operatorname{min}\{J_1,J_2,2\}}\nonumber\\
	&\quad\times\sum^{\operatorname{min}\{J_2,J_3\}}_{d'=0}(1+\delta_{d0})^2
	\gamma_{J_3}\alpha^{J_1J_2}_{d}\epsilon^{J_3J_2}_{d'}K_{J_1d}\nonumber\\
	&\quad\times\left[\beta^{J_1}_{d}D^{J_3*}_{d'0}(\theta'',\phi'')D^{J_2*}_{dd'}(\theta',\phi')\right.\nonumber\\
		&\quad\left.+(-1)^{J_1}\beta^{J_1*}_dD^{J_3}_{d'0}(\theta'',\phi'')D^{J_2}_{dd'}(\theta',\phi')\right]
\end{align}
Using the orthogonaity of the Wigner $D$-functions, the coefficients for all possible values of $J_1$, $J_2$, $J_3$, $d$ and $M$ in \eqref{eq:Case6} can be obtained from
\begin{flalign}\label{eq:coefficients 2}
&\gamma_{J_3}\epsilon^{J_3J_2}_d\alpha^{J_1J_2}_M K_{J_1M}[1+\delta_{M0}]^2[1+\delta_{d0}\delta_{M0}]\nonumber\\
	&\quad\times[\beta^{J_1}_M+(-1)^{J_1}\beta^{J_1*}_{M}]\nonumber\\
	&=16(2J_2+1)(2J_3+1)(2-\delta_{M0})\int\widetilde{\widetilde{W}}(\theta',\phi';\theta'',\phi'')\nonumber\\
		&\quad\times[D^{J_3}_{d0}D^{J_2}_{Md}+D^{J_3*}_{d0}D^{J_2*}_{Md}]d\Omega'd\Omega''\text{.}
\end{flalign}
From \eqref{eq:coefficients 2}, we can determine the relative magnitudes as well as the relative phases of all the $E$ and $B$ helicity amplitudes by measuring the simultaneous angular distribution of $\gamma_2$ and $e^-$. Moreover, the relative phases of the $A$ helicity amplitudes can also be obtained.

\section{Concluding remarks}
We have derived a model-independent expression for the combined angular distribution of the final electron and the two gamma photons in the cascade process, $\bar{p}p\rightarrow{}^3D_3\rightarrow{}^3P_2+\gamma_1\rightarrow\psi+\gamma_2+\gamma_1\rightarrow e^++e^-+\gamma_2+\gamma_1$, when $\bar{p}$ and $p$ are arbitrarily polarized. Our expression is based only on the general principles of quantum mechanics and the symmetry of the problem. We have also derived the partially integrated angular distribution functions which give the angular distributions of $\gamma_1$, $\gamma_2$,  and $e^-$ alone and of ($\gamma_1$,$\gamma_2$), ($\gamma_1$,$e^-$) and ($\gamma_2$,$e^-$). Once these angular distributions are experimentally measured, our expressions can be used to extract all the independent helicity amplitudes in the two radiative decay processes ${}^3D_3\rightarrow{}^3P_2+\gamma_1$ and ${}^3P_2\rightarrow\psi+\gamma_2$. In fact, the analysis of the angular correlations in the final decay products will serve to verify the presence of the intermediate ${}^3D_3$ charmonium state and its $J^{PC}$ quantum numbers in the cascade process. The experimentally determined values of the helicity amplitudes can then be compared with the predictions of various dynamical models.

The great advantage of studying polarized $\bar{p}p$ collisions is that one can obtain not only the relative magnitudes of the helicity amplitudes but also both the cosines and the sines of the relative phases of the helicity amplitudes from the measurement of the simultaneous angular distributions of two particles. This is important because the helicity amplitudes are in general complex. Therefore, we can get complete information on all the helicity amplitudes in the process ${}^3D_3\rightarrow{}^3P_2+\gamma_1$ from the simultaneous angular distribution of $\gamma_1$ and $\gamma_2$ and also in the process ${}^3P_2\rightarrow\psi+\gamma_2$ from the simultaneous angular distribution of $\gamma_2$ and $e^-$, when both $\bar{p}$ and $p$ are polarized with both transverse and longitudinal polarization vector components in their respective frames. Moreover, we can also obtain the relative magnitude and the relative phase of the helicity amplitudes in the process $\bar{p}p\rightarrow{}^3D_3$. Polarizations of both $\bar{p}$ and $p$ are necessary to get all this information. Alternatively, one can also consider the polarizations of the final decay products $\gamma_1$, $\gamma_2$ and $e^-$ \cite{mok and chow,ref 16}.

We should also emphasize that the angular distributions alone will not give the absolute strengths of the helicity amplitudes. We get the magnitudes of all the helicity amplitudes only with the arbitrary normalization convention of \eqref{eq: normalizations 1} and \eqref{eq: normalizations 2}. In order to get the true absolute values which are physically significant one has to measure the branching ratios of each of the above processes and the parent particle's lifetime or decay width. The measurement of the angular distributions alone will only give the relative magnitudes and the relative phases of the helicity amplitudes in each radiative decay process.

Both the theorists and the experimentalists would like to express their results in terms of the multipole amplitudes in the radiative transitions ${}^3D_3\rightarrow{}^3P_2+\gamma_1$ and ${}^3P_2\rightarrow\psi+\gamma_2$. The relationship between the helicity and the multipole amplitudes are given by the orthogonal transformations \cite{ref 17,ref 18}
\begin{align}
\label{eq:orthogonal transformations 1}
&A_i=\sum^5_{k=1}a_k\sqrt{\frac{2k+1}{5}}\langle k,-1;3,(i-1)|2,(i-2)\rangle\text{;}\\
&i=0\text{, }1\text{, }2\text{, }3\text{, }4\text{, }\nonumber\\
\label{eq:orthogonal transformations 2}
&E_j=\sum^3_{k=1}e_k\sqrt{\frac{2k+1}{5}}\langle k,1;1,(j-1)|2,j\rangle\text{;}\\
&j=0\text{, }1\text{, }2\text{, }\nonumber
\end{align}
where $a_k$ and $e_k$ are the radiative multipole amplitudes in ${}^3D_3\rightarrow{}^3P_2+\gamma_1$ and ${}^3P_2\rightarrow\psi+\gamma_2$, respectively. Since the transformations of \eqref{eq:orthogonal transformations 1} and \eqref{eq:orthogonal transformations 2} are orthogonal,
\begin{align}
\sum^4_{i=0}|A_i|^2=\sum^5_{k=1}|a_k|^2=1\text{,}\quad\sum^2_{j=0}|E_j|^2=\sum^3_{k=1}|e_k|^2=1\text{.}
\end{align}

\appendix
\section{Expressions of Coefficients}\label{appendix}
\subsection*{Expressions of $\beta^{J_1}_M$}
\begin{flalign}
\beta^0_0=&|B_0|^2(P_-+P_A)+|B_1|^2P_+\\
\beta^1_0=&\frac{1}{2}|B_1|^2(P_{1z}+P_{2z})\\
\beta^2_0=&-\frac{2}{\sqrt{3}}\Big[|B_0|^2(P_-+P_A)+\frac{3}{4}|B_1|^2P_+\Big]\\
\beta^3_0=&-\sqrt{42}|B_1|^2(P_{1z}+P_{2z})\\
\beta^4_0=&3\sqrt{\frac{2}{11}}\Big[|B_0|^2(P_-+P_A)+\frac{1}{6}|B_1|^2P_+\Big]\\
\beta^5_0=&\frac{5}{2\sqrt{3}}|B_1|^2(P_{1z}+P_{2z})\\
\beta^6_0=&-10\sqrt{\frac{1}{33}}\Big[|B_0|^2(P_-+P_A)-\frac{3}{4}|B_1|^2P_+\Big]\\
\beta^1_1=&-\sqrt{\frac{3}{2}}\Big[\text{Re}(B_0B_1^*)(P_{1x}+P_{2x})-\text{Im}(B_0B_1^*)P_D\Big]\nonumber\\
	&+\sqrt{\frac{3}{2}}\operatorname{i}\Big[\text{Re}(B_0B_1^*)(P_{1y}+P_{2y})\nonumber\\
	&+\text{Im}(B_0B_1^*)P_E\Big]\\
\beta^2_1=&\sqrt{\frac{1}{6}}\Big[-\text{Re}(B_0B_1^*)P_{E}+\text{Im}(B_0B_1^*)(P_{1y}+P_{2y})\Big]\nonumber\\
	&-\sqrt{\frac{1}{6}}\operatorname{i}\Big[-\text{Re}(B_0B_1^*)P_{D}\nonumber\\
	&-\text{Im}(B_0B_1^*)(P_{1x}+P_{2x})\Big]\\
\beta^3_1=&\sqrt{\frac{7}{6}}\Big[\text{Re}(B_0B_1^*)(P_{1x}+P_{2x})-\text{Im}(B_0B_1^*)P_D\Big]\nonumber\\
	&-\sqrt{\frac{7}{6}}\operatorname{i}\Big[\text{Re}(B_0B_1^*)(P_{1y}+P_{2y})\nonumber\\
	&+\text{Im}(B_0B_1^*)P_E\Big]\\
\beta^4_1=&-\sqrt{\frac{15}{22}}\Big[\text{Re}(B_0B_1^*)P_{E}-\text{Im}(B_0B_1^*)(P_{1y}+P_{2y})\Big]\nonumber\\
	&+\sqrt{\frac{15}{22}}\operatorname{i}\Big[-\text{Re}(B_0B_1^*)P_{D}\nonumber\\
	&-\text{Im}(B_0B_1^*)(P_{1x}+P_{2x})\Big]\\
\beta^5_1=&-\sqrt{\frac{5}{6}}\Big[\text{Re}(B_0B_1^*)(P_{1x}+P_{2x})-\text{Im}(B_0B_1^*)P_D\Big]\nonumber\\
	&+\sqrt{\frac{5}{6}}\operatorname{i}\Big[\text{Re}(B_0B_1^*)(P_{1y}+P_{2y})\nonumber\\
	&+\text{Im}(B_0B_1^*)P_E\Big]\\
\beta^6_1=&\frac{5}{2}\sqrt{\frac{14}{33}}\Big[-\text{Re}(B_0B_1^*)P_{E}+\text{Im}(B_0B_1^*)(P_{1y}+P_{2y})\Big]\nonumber\\
	&+\frac{5}{2}\sqrt{\frac{14}{33}}\operatorname{i}\Big[\text{Re}(B_0B_1^*)P_{D}\nonumber\\
	&+\text{Im}(B_0B_1^*)(P_{1x}+P_{2x})\Big]\\
\beta^2_2=&\frac{1}{\sqrt{2}}|B_1|^2(P_B-\operatorname{i}P_C)\\
\beta^4_2=&-\sqrt{\frac{5}{11}}|B_1|^2(P_B-\operatorname{i}P_C)\\
\beta^6_2=&\frac{1}{2}\sqrt{\frac{35}{11}}|B_1|^2(P_B-\operatorname{i}P_C)
\end{flalign}

\subsection*{Expressions of $\alpha^{J_1J_2}_d$}
\begin{flalign}
\alpha_0^{00}=&|A_0|^2+|A_1|^2+|A_2|^2+|A_3|^2+|A_4|^2=1\\
\alpha_0^{02}=&\sqrt{\frac{10}{7}}\Bigg(|A_0|^2-\frac{1}{2}|A_1|^2-|A_2|^2-\frac{1}{2}|A_3|^2+|A_4|^2\Bigg)\\
\alpha_0^{04}=&\sqrt{\frac{1}{14}}(|A_0|^2-4|A_1|^2+6|A_2|^2-4|A_3|^2+|A_4|^2)\\
\alpha_0^{11}=&\frac{3}{\sqrt{2}}\Bigg(|A_0|^2+\frac{1}{3}|A_1|^2+\frac{1}{3}|A_4|^2\Bigg)\\
\alpha_1^{11}=&\sqrt{3}\Big[\text{Re}(A_1A_0^*)+\sqrt{\frac{5}{2}}\text{Re}(A_2A_1^*)+\sqrt{3}\text{Re}(A_3A_2^*)\nonumber\\
	&+\sqrt{2}\text{Re}(A_4A_3^*)\Big]\\
\alpha_1^{12}=&-3\sqrt{\frac{5}{7}}\operatorname{i}\Big[\text{Im}(A_1A_0^*)+\frac{1}{3}\sqrt{\frac{5}{2}}\text{Im}(A_2A_1^*)\nonumber\\
	&-\frac{1}{\sqrt{3}}\text{Im}(A_3A_2^*)-\sqrt{2}\text{Im}(A_4A_3^*)\Big]\\
\alpha_0^{13}=&\frac{3}{2\sqrt{2}}\Bigg(|A_0|^2-\frac{4}{3}|A_1|^2+\frac{1}{3}|A_4|^2\Bigg)\\
\alpha_1^{13}=&\frac{3}{\sqrt{2}}\Big[\text{Re}(A_1A_0^*)-\frac{\sqrt{10}}{3}\text{Re}(A_2A_1^*)-\frac{2}{\sqrt{3}}\text{Re}(A_3A_2^*)\nonumber\\
	&+\sqrt{2}\text{Re}(A_4A_3^*)\Big]\\
\alpha_1^{14}=&-3\sqrt{\frac{5}{42}}\operatorname{i}\Big[\text{Im}(A_1A_0^*)-\sqrt{10}\text{Im}(A_2A_1^*)\nonumber\\
	&+2\sqrt{3}\text{Im}(A_3A_2^*)-\sqrt{2}\text{Im}(A_4A_3^*)\Big]\\
\alpha_0^{20}=&\frac{5}{2\sqrt{3}}\Bigg(|A_0|^2-\frac{3}{5}|A_2|^2-\frac{4}{5}|A_3|^2-\frac{3}{5}|A_4|^2\Bigg)\\
\alpha_1^{21}=&-\frac{5}{\sqrt{3}}\operatorname{i}\Big[\text{Im}(A_1A_0^*)+\frac{3}{\sqrt{10}}\text{Im}(A_2A_1^*)\nonumber\\
	&+\frac{\sqrt{3}}{5}\text{Im}(A_3A_2^*)-\frac{\sqrt{2}}{5}\text{Im}(A_4A_3^*)\Big]\\
\alpha_0^{22}=&5\sqrt{\frac{5}{42}}\Big(|A_0|^2+\frac{3}{5}|A_2|^2+\frac{2}{5}|A_3|^2-\frac{3}{5}|A_4|^2\Big)\\
\alpha_1^{22}=&5\sqrt{\frac{5}{7}}\Big[\text{Re}(A_1A_0^*)+\frac{1}{\sqrt{10}}\text{Re}(A_2A_1^*)\nonumber\\
	&-\frac{1}{5\sqrt{3}}\text{Re}(A_3A_2^*)+\frac{\sqrt{2}}{5}\text{Re}(A_4A_3^*)\Big]\\
\alpha_2^{22}=&\frac{10}{\sqrt{21}}\Big[\text{Re}(A_2A_0^*)+\sqrt{3}\text{Re}(A_3A_1^*)\nonumber\\
	&+2\sqrt{\frac{3}{5}}\text{Re}(A_4A_2^*)\Big]\\
\alpha_1^{23}=&-\frac{5}{\sqrt{2}}\operatorname{i}\Big[\text{Im}(A_1A_0^*)-\sqrt{\frac{2}{5}}\text{Im}(A_2A_1^*)\nonumber\\
	&-\frac{2}{5\sqrt{3}}\text{Im}(A_3A_2^*)-\frac{\sqrt{2}}{5}\text{Im}(A_4A_3^*)\Big]\\
\alpha_2^{23}=&-\frac{5}{\sqrt{3}}\operatorname{i}\Big[\text{Im}(A_2A_0^*)-2\sqrt{\frac{3}{5}}\text{Im}(A_4A_2^*)\Big]\\
\alpha_0^{24}=&\frac{5}{2\sqrt{42}}\Bigg(|A_0|^2-\frac{18}{5}|A_2|^2+\frac{16}{5}|A_3|^2-\frac{3}{5}|A_4|^2\Bigg)\\
\alpha_1^{24}=&5\sqrt{\frac{5}{42}}\Big[\text{Re}(A_1A_0^*)-3\sqrt{\frac{2}{5}}\text{Re}(A_2A_1^*)\nonumber\\
	&+\frac{2\sqrt{3}}{5}\text{Re}(A_3A_2^*)+\frac{\sqrt{2}}{5}\text{Re}(A_4A_3^*)\Big]\\
\alpha_2^{24}=&\frac{5}{\sqrt{7}}\Big[\text{Re}(A_2A_0^*)-\frac{4}{\sqrt{3}}\text{Re}(A_3A_1^*)\nonumber\\
	&+2\sqrt{\frac{3}{5}}\text{Re}(A_4A_2^*)\Big]\\
\alpha_0^{31}=&\sqrt{\frac{7}{3}}\Bigg(|A_0|^2-\frac{1}{2}|A_1|^2-|A_4|^2\Bigg)\\
\alpha_1^{31}=&2\sqrt{\frac{7}{3}}\Big[\text{Re}(A_1A_0^*)-\frac{\sqrt{3}}{2}\text{Re}(A_3A_2^*)\nonumber\\
	&-\sqrt{2}\text{Re}(A_4A_3^*)\Big]\\
\alpha_1^{32}=&-2\sqrt{5}\operatorname{i}\Big[\text{Im}(A_1A_0^*)+\frac{1}{2\sqrt{3}}\text{Im}(A_3A_2^*)\nonumber\\
	&+\frac{1}{\sqrt{2}}\text{Im}(A_4A_3^*)\Big]\\
\alpha_2^{32}=&-2\sqrt{\frac{10}{3}}\operatorname{i}\Big[\text{Im}(A_2A_0^*)+\frac{\sqrt{3}}{2}\text{Im}(A_3A_1^*)\Big]\\
\alpha_0^{33}=&\sqrt{\frac{7}{12}}(|A_0|^2+2|A_1|^2-|A_4|^2)\\
\alpha_1^{33}=&\sqrt{14}\Big[\text{Re}(A_1A_0^*)+\frac{1}{\sqrt{3}}\text{Re}(A_3A_2^*)-\frac{1}{\sqrt{2}}\text{Re}(A_4A_3^*)\Big]\\
\alpha_2^{33}=&\sqrt{\frac{70}{3}}\text{Re}(A_2A_0^*)\\
\alpha_3^{33}=&\sqrt{\frac{35}{3}}\Big[\text{Re}(A_3A_0^*)+\sqrt{2}\text{Re}(A_4A_1^*)\Big]\\
\alpha_1^{34}=&-\sqrt{\frac{10}{3}}\operatorname{i}\Big[\text{Im}(A_1A_0^*)-\sqrt{3}\text{Im}(A_3A_2^*)\nonumber\\
	&+\frac{1}{\sqrt{2}}\text{Im}(A_4A_3^*)\Big]\\
\alpha_2^{34}=&-\sqrt{10}\operatorname{i}\Big[\text{Im}(A_2A_0^*)-\frac{2}{\sqrt{3}}\text{Im}(A_3A_1^*)\Big]\\
\alpha_3^{34}=&-\sqrt{\frac{35}{3}}\operatorname{i}\Big[\text{Im}(A_3A_0^*)-\sqrt{2}\text{Im}(A_4A_1^*)\Big]\\
\alpha_0^{40}=&\frac{3}{\sqrt{22}}\Bigg(|A_0|^2-\frac{7}{3}|A_1|^2\nonumber\\
	&+\frac{1}{3}|A_2|^2+2|A_3|^2+\frac{1}{3}|A_4|^2\Bigg)\\
\alpha_1^{41}=&-2\sqrt{\frac{15}{11}}\operatorname{i}\Big[\text{Im}(A_1A_0^*)-2\sqrt{\frac{2}{5}}\text{Im}(A_2A_1^*)\nonumber\\
	&-\frac{\sqrt{3}}{2}\text{Im}(A_3A_2^*)
	+\frac{1}{\sqrt{2}}\text{Im}(A_4A_3^*)\Big]\\
\alpha_0^{42}=&3\sqrt{\frac{5}{77}}\Bigg(|A_0|^2+\frac{7}{6}|A_1|^2\nonumber\\
	&-\frac{1}{3}|A_2|^2-|A_3|^2+\frac{1}{3}|A_4|^2\Bigg)\\
\alpha_1^{42}=&\frac{30}{\sqrt{77}}\Big[\text{Re}(A_1A_0^*)-\frac{2}{3}\sqrt{\frac{2}{5}}\text{Re}(A_2A_1^*)\nonumber\\
	&+\frac{1}{2\sqrt{3}}\text{Re}(A_3A_2^*)-\frac{1}{\sqrt{2}}\text{Re}(A_4A_3^*)\Big]\\
\alpha_2^{42}=&6\sqrt{\frac{30}{77}}\Big[\text{Re}(A_2A_0^*)-\frac{1}{2\sqrt{3}}\text{Re}(A_3A_1^*)\nonumber\\
	&-\frac{2}{3}\sqrt{\frac{5}{3}}\text{Re}(A_4A_2^*)\Big]\\
\alpha_1^{43}=&-3\sqrt{\frac{10}{11}}\operatorname{i}\Big[\text{Im}(A_1A_0^*)+\frac{4}{3}\sqrt{\frac{2}{5}}\text{Im}(A_2A_1^*)\nonumber\\
	&+\frac{1}{\sqrt{3}}\text{Im}(A_3A_2^*)+\frac{1}{\sqrt{2}}\text{Im}(A_4A_3^*)\Big]\\
\alpha_2^{43}=&-3\sqrt{\frac{30}{11}}\operatorname{i}\Big[\text{Im}(A_2A_0^*)+\frac{2}{3}\sqrt{\frac{5}{3}}\text{Im}(A_4A_2^*)\Big]\\
\alpha_3^{43}=&-3\sqrt{\frac{35}{11}}\operatorname{i}\Big[\text{Im}(A_3A_0^*)+\frac{\sqrt{2}}{3}\operatorname{Im}(A_4A_1^*)\Big]\\
\alpha_0^{44}=&\frac{3}{2\sqrt{77}}\Bigg(|A_0|^2+\frac{28}{3}|A_1|^2\nonumber\\
	&+2|A_2|^2-8|A_3|^2+\frac{1}{3}|A_4|^2\Bigg)\\
\alpha_1^{44}=&5\sqrt{\frac{6}{77}}\Big[\text{Re}(A_1A_0^*)+4\sqrt{\frac{2}{5}}\text{Re}(A_2A_1^*)\nonumber\\
	&-\sqrt{3}\text{Re}(A_3A_2^*)-\frac{1}{\sqrt{2}}\text{Re}(A_4A_3^*)\Big]\\
\alpha_2^{44}=&9\sqrt{\frac{10}{77}}\Big[\text{Re}(A_2A_0^*)+\frac{2}{3\sqrt{3}}\text{Re}(A_3A_1^*)\nonumber\\
	&-\frac{2}{3}\sqrt{\frac{5}{3}}\text{Re}(A_4A_2^*)\Big]\\
\alpha_3^{44}=&3\sqrt{\frac{35}{11}}\Big[\text{Re}(A_3A_0^*)-\frac{\sqrt{2}}{3}\operatorname{Re}(A_4A_1^*)\Big]\\
\alpha_4^{44}=&2\sqrt{\frac{105}{11}}\text{Re}(A_4A_0^*)\\
\alpha_0^{51}=&\frac{1}{\sqrt{6}}(|A_0|^2-2|A_1|^2+5|A_4|^2)\\
\alpha_1^{51}=&\sqrt{\frac{5}{3}}\Big[\text{Re}(A_1A_0^*)-\frac{9}{\sqrt{10}}\text{Re}(A_2A_1^*)+\sqrt{3}\text{Re}(A_3A_2^*)\nonumber\\
	&+\sqrt{2}\text{Re}(A_4A_3^*)\Big]\\
\alpha_1^{52}=&-\frac{5}{\sqrt{7}}\operatorname{i}\Big[\text{Im}(A_1A_0^*)-\frac{3}{\sqrt{10}}\text{Im}(A_2A_1^*)\nonumber\\
	&-\frac{1}{\sqrt{3}}\text{Im}(A_3A_2^*)-\sqrt{2}\text{Im}(A_4A_3^*)\Big]\\
\alpha_2^{52}=&-2\sqrt{\frac{5}{3}}\operatorname{i}\Big[\text{Im}(A_2A_0^*)-\sqrt{3}\text{Im}(A_3A_1^*)\Big]\\
\alpha_0^{53}=&\frac{1}{2\sqrt{6}}(|A_0|^2+8|A_1|^2+5|A_4|^2)\\
\alpha_1^{53}=&\sqrt{\frac{5}{2}}\Big[\text{Re}(A_1A_0^*)+3\sqrt{\frac{2}{5}}\text{Re}(A_2A_1^*)\nonumber\\
	&-\frac{2}{\sqrt{3}}\text{Re}(A_3A_2^*)+\sqrt{2}\text{Re}(A_4A_3^*)\Big]\\
\alpha_2^{53}=&\sqrt{\frac{35}{3}}\text{Re}(A_2A_0^*)\\
\alpha_3^{53}=&\sqrt{\frac{70}{3}}\Big[\text{Re}(A_3A_0^*)-\frac{1}{\sqrt{2}}\text{Re}(A_4A_1^*)\Big]\\
\alpha_1^{54}=&-\frac{5}{\sqrt{42}}\operatorname{i}\Big[\text{Im}(A_1A_0^*)+9\sqrt{\frac{2}{5}}\text{Im}(A_2A_1^*)\nonumber\\
	&+2\sqrt{3}\text{Im}(A_3A_2^*)-\sqrt{2}\text{Im}(A_4A_3^*)\Big]\\
\alpha_2^{54}=&-\sqrt{5}\operatorname{i}\Big[\text{Im}(A_2A_0^*)+\frac{4}{\sqrt{3}}\text{Im}(A_3A_1^*)\Big]\\
\alpha_3^{54}=&-2\sqrt{\frac{35}{6}}\operatorname{i}\Big[\text{Im}(A_3A_0^*)+\frac{1}{\sqrt{2}}\text{Im}(A_4A_1^*)\Big]\\
\alpha_4^{54}=&-\sqrt{70}\operatorname{i}\text{Im}(A_4A_0^*)\\
\alpha_0^{60}=&\frac{1}{2\sqrt{33}}(|A_0|^2-6|A_1|^2\nonumber\\
	&+15|A_2|^2-20|A_3|^2+15|A_4|^2)\\
\alpha_1^{61}=&-\sqrt{\frac{7}{33}}\operatorname{i}\Big[\text{Im}(A_1A_0^*)-3\sqrt{\frac{5}{2}}\text{Im}(A_2A_1^*)\nonumber\\
	&+5\sqrt{3}\text{Im}(A_3A_2^*)-\frac{10}{\sqrt{2}}\text{Im}(A_4A_3^*)\Big]\\
\alpha_0^{62}=&\sqrt{\frac{5}{462}}(|A_0|^2+3|A_1|^2\nonumber\\
	&-15|A_2|^2+10|A_3|^2+15|A_4|^2)\\
\alpha_1^{62}=&\sqrt{\frac{5}{11}}\Big[\text{Re}(A_1A_0^*)-\sqrt{\frac{5}{2}}\text{Re}(A_2A_1^*)\nonumber\\
	&-\frac{5}{\sqrt{3}}\text{Re}(A_3A_2^*)
	+\frac{10}{\sqrt{2}}\text{Re}(A_4A_3^*)\Big]\\
\alpha_2^{62}=&2\sqrt{\frac{10}{33}}\Big[\text{Re}(A_2A_0^*)-2\sqrt{3}\text{Re}(A_3A_1^*)\nonumber\\
	&+\sqrt{15}\text{Re}(A_4A_2^*)\Big]\\
\alpha_1^{63}=&-\sqrt{\frac{7}{22}}\operatorname{i}\Big[\text{Im}(A_1A_0^*)+\sqrt{10}\text{Im}(A_2A_1^*)\nonumber\\
	&-\frac{10}{\sqrt{3}}\text{Im}(A_3A_2^*)-5\sqrt{2}\text{Im}(A_4A_3^*)\Big]\\
\alpha_2^{63}=&-\sqrt{\frac{70}{33}}\operatorname{i}\Big[\text{Im}(A_2A_0^*)-\sqrt{15}\text{Im}(A_4A_2^*)\Big]\\
\alpha_3^{63}=&-\sqrt{\frac{70}{11}}\operatorname{i}\Big[\text{Im}(A_3A_0^*)-\frac{3}{\sqrt{2}}\text{Im}(A_4A_1^*)\Big]\\
\alpha_0^{64}=&\frac{1}{2\sqrt{462}}(|A_0|^2+24|A_1|^2\nonumber\\
	&+90|A_2|^2+80|A_3|^2+15|A_4|^2)\\
\alpha_1^{64}=&\sqrt{\frac{5}{66}}\Big[\text{Re}(A_1A_0^*)+3\sqrt{10}\text{Re}(A_2A_1^*)\nonumber\\
	&+10\sqrt{3}\text{Re}(A_3A_2^*)+5\sqrt{2}\text{Re}(A_4A_3^*)\Big]\\
\alpha_2^{64}=&\sqrt{\frac{10}{11}}\Big[\text{Re}(A_2A_0^*)+\frac{8}{\sqrt{3}}\text{Re}(A_3A_1^*)\nonumber\\
	&+\sqrt{15}\text{Re}(A_4A_2^*)\Big]\\
\alpha_3^{64}=&\sqrt{\frac{70}{11}}\Big[\text{Re}(A_3A_0^*)+\frac{3}{\sqrt{2}}\text{Re}(A_4A_1^*)\Big]\\
\alpha_4^{64}=&5\sqrt{\frac{14}{11}}\text{Re}(A_4A_0^*)
\end{flalign}

\subsection*{Expressions of $\epsilon_{d'}^{J_3J_2}$}
\begin{flalign} 
\epsilon_0^{00}=&|E_0|^2+|E_1|^2+|E_2|^2=1\\
\epsilon_0^{02}=&-\sqrt{\frac{10}{7}}\Bigg(|E_0|^2+\frac{1}{2}|E_1|^2-|E_2|^2\Bigg)\\
\epsilon_0^{04}=&3\sqrt{\frac{2}{7}}\Bigg(|E_0|^2-\frac{2}{3}|E_1|^2+\frac{1}{6}|E_2|^2\Bigg)\\
\epsilon_0^{20}=&\frac{1}{\sqrt{2}}(|E_0|^2-2|E_1|^2+|E_2|^2)\\
\epsilon_1^{21}=&-3\operatorname{i}\Big[\text{Im}(E_1E_0^*)-\sqrt{\frac{2}{3}}\text{Im}(E_2E_1^*)\Big]\\
\epsilon_0^{22}=&-\sqrt{\frac{5}{7}}(|E_0|^2-|E_1|^2-|E_2|^2)\\
\epsilon_1^{22}=&-\sqrt{\frac{15}{7}}\Big[\text{Re}(E_1E_0^*)-\sqrt{6}\text{Re}(E_2E_1^*)\Big]\\
\epsilon_2^{22}=&2\sqrt{\frac{30}{7}}\text{Re}(E_2E_0^*)\\
\epsilon_1^{23}=&\sqrt{6}\operatorname{i}\Big[\text{Im}(E_1E_0^*)+\sqrt{\frac{3}{2}}\text{Im}(E_2E_1^*)\Big]\\
\epsilon_2^{23}=&\sqrt{30}\operatorname{i}\text{Im}(E_2E_0^*)\\
\epsilon_0^{24}=&\frac{3}{\sqrt{7}}\Bigg(|E_0|^2+\frac{4}{3}|E_1|^2+\frac{1}{6}|E_2|^2\Bigg)\\
\epsilon_1^{24}=&\sqrt{\frac{90}{7}}\Big[\text{Re}(E_1E_0^*)+\sqrt{\frac{1}{6}}\text{Re}(E_2E_1^*)\Big]\\
\epsilon_2^{24}=&\sqrt{\frac{90}{7}}\text{Re}(E_2E_0^*)
\end{flalign}

\subsection*{Expressions of $\gamma_{J_3}$}
\begin{flalign}
\gamma_0=&|C_0|^2+|C_1|^2=1\\
\gamma_2=&\sqrt{2}\Bigg(|C_0|^2-\frac{1}{2}|C_1|^2\Bigg)\cong\sqrt{\frac{1}{2}}
\end{flalign}

\end{document}